\documentclass[%
 preprint,
groupedaddress,
superscriptaddress,
 amsmath,amssymb,
 aps,
 prx,
 showpacs
]{revtex4-2}

\usepackage{graphicx}
\usepackage{makecell,booktabs}
\usepackage{dcolumn}
\usepackage{bm}
\usepackage{hyperref}
\hypersetup{
    colorlinks=true,
    urlcolor= blue,
    citecolor=blue,
linkcolor= blue}
\usepackage{xcolor}

\newcommand{\NiFe}{$\mathrm{Ni_{0.8}Fe_{0.2}}$}
\begin{document}


\title{Giant spin torque efficiency in naturally oxidized polycrystalline TaAs thin films}

\author{Wilson Yanez}
\author{Yongxi Ou}
\author{Run Xiao}

 \affiliation{Department of Physics, Pennsylvania State University, University Park, Pennsylvania 16802, USA}

\author{Supriya Ghosh}
\affiliation{%
Department of Chemical Engineering and Materials Science, University of Minnesota, Minneapolis, Minnesota 55455, USA
}%
\author{Jyotirmay Dwivedi}
\author{Emma Steinebronn}
 \affiliation{Department of Physics, Pennsylvania State University, University Park, Pennsylvania 16802, USA}
 \author{Anthony Richardella}
 \affiliation{Department of Physics, Pennsylvania State University, University Park, Pennsylvania 16802, USA}

\author{K. Andre Mkhoyan}
\affiliation{%
Department of Chemical Engineering and Materials Science, University of Minnesota, Minneapolis, Minnesota 55455, USA
}%
\author{Nitin Samarth}
\email{nsamarth@psu.edu}
\affiliation{Department of Physics, Pennsylvania State University, University Park, Pennsylvania 16802, USA}

\date{\today}

\begin{abstract}
We report the measurement of efficient charge-to-spin conversion at room temperature in Weyl semimetal/ferromagnet heterostructures with both oxidized and pristine interfaces. Polycrystalline films of the Weyl semimetal, TaAs, are grown by molecular beam epitaxy on (001) GaAs and interfaced with a metallic ferromagnet (\NiFe). Spin torque ferromagnetic resonance (ST-FMR) measurements in samples with an oxidized interface yield a spin torque efficiency as large as $\xi_{\mathrm{FMR}}=0.45\pm 0.25$ for a 8 nm \NiFe~layer thickness. By studying ST-FMR in these samples with varying \NiFe~layer thickness, we extract a damping-like torque efficiency as high as  $\xi_{\mathrm{DL}}=1.36\pm 0.66$. In samples with a pristine (unoxidized) interface, the spin torque efficiency has opposite sign to that observed in oxidized samples ($\xi_{\mathrm{FMR}}=-0.27\pm 0.14$ for a 5 nm \NiFe~layer thickness). We also find a lower bound on the spin Hall conductivity ($424 \pm 110 \frac{\hbar}{e}$ S/cm) which is surprisingly consistent with theoretical predictions for the single crystal Weyl semimetal state of TaAs. 

\end{abstract}

\maketitle



The electrical manipulation of the spin degree of freedom in a material is a fundamental scientific aim that has a direct impact on nonvolatile, low dissipation magnetic memory applications. In this context, heavy metals and topological insulators have proven to be attractive platforms to study charge-spin interconversion due to the presence of significant spin Hall and Rashba-Edelstein effects \cite{doi:10.1126/science.1218197GiantTa,PhysRevLett.106.036601LiuSTFMR,PhysRevB.92.064426CFPaiSTFMR,Mellnik2014,PhysRevLett.117.076601hailong,PhysRevResearch.1.012014,Wang2017TIswitch,PhysRevLett.123.207205,Kondou2016}. More recently, Dirac and Weyl semimetals have started attracting interest for spintronics because of the potential for combining topological aspects of the band structure with increased electrical conductivity \cite{MacNeill2017WTe2,PhysRevB.96.054450Wte2,Shi2019,PhysRevApplied.16.054031yanez,https://doi.org/10.1002/adma.202000513xu,PhysRevLett.124.116802FermiArcSpinCd3as2,doi:10.1021/acsnano.1c00154cd3as2spin}. In particular, TaAs is an archetypal topological Weyl semimetal of contemporary interest \cite{XuTaAsArpesScience} that has spin polarized surface states \cite{XuPhysRevLett.116.096801}. Although it is theoretically predicted to have a giant intrinsic spin Hall conductivity \cite{PhysRevLett.117.146403SHCtheoryBY}, this has yet to be demonstrated experimentally. Additionally, the role of natural oxidation has usually been neglected in topological materials, even though it has been shown that it can enhance the charge-spin interconversion efficiency in heavy metals and Dirac semimetals \cite{An2016CuOx,Tsai2018BiOx,PhysRevApplied.16.054031yanez}. Several mechanisms have been proposed to explain this phenomenon, such as the presence of Rashba spin textures at the interface  \cite{PhysRevLett.108.186802LAO1,Noel2020LAO2,PhysRevResearch.2.012014LaO3} or a theoretically predicted giant orbital Hall effect that can be further enhanced in oxide interfaces \cite{PhysRevB.77.165117orbitalhall,PhysRevB.103.L121113OrbitalOxide}.

In this paper, we report measurements of the charge-spin interconversion phenomenon in polycrystalline thin films of TaAs interfaced with a metallic ferromagnet (\NiFe) with and without natural surface oxidation. We show that at room temperature, the spin torque efficiency in oxidized TaAs films can be as large as $\xi_{\mathrm{FMR}}=0.45\pm 0.25$. We also present measurements of a completely \textit{in vacuo} grown TaAs/$\mathrm{Ni_{80}Fe_{20}}$ heterostructure, obtaining an ST-FMR efficiency $\xi_{\mathrm{FMR}}=-0.27\pm 0.14$ with a spin Hall conductivity that agrees quantitatively with prior \textit{ab initio} calculations for the Weyl semimetal phase of TaAs \cite{PhysRevLett.117.146403SHCtheoryBY}. This is surprising since the films measured here are polycrystalline. Notably, the sign of the symmetric component of the ST-FMR spectra changes with oxidation indicating that the sign of the spin torque efficiency is opposite for pristine versus oxidized samples.  We then present a systematic study of the spin torque efficiency in oxidized TaAs samples as a function of the ferromagnet layer thickness and extract a damping-like torque efficiency as high as  $\xi_{\mathrm{DL}}=1.36\pm 0.66$. We also observe the presence of a significant field-like torque. Finally, we show that the sign change of the the spin torque efficiency also occurs in simple metal Ta/\NiFe~heterostructures when comparing oxidized and pristine interfaces.  

We first discuss the synthesis of polycrystalline TaAs thin films using molecular beam epitaxy (MBE). TaAs crystallizes in a body-centered tetragonal structure ($\mathrm{I4_{1}md}$ space group) with a lattice constant of $a = 0.3437$ nm and $c=1.1656$ nm \cite{XuTaAsArpesScience,XuPhysRevLett.116.096801}. This makes it challenging to find an appropriate substrate for its thin film epitaxial growth given that most commonly available semiconductors do not have lattice constants in this range \cite{doi:10.1063/1.1368156bandgap}. We note that MBE growth of thin films of the related Weyl semimetals, NbP and TaP, has been reported using a thin Nb buffer layer on (001) MgO \cite{Bedoya-Pinto_ACSNano}. We tried different substrates and find that GaAs (001) has the correct surface chemistry to nucleate TaAs growth. We carried out the synthesis in a VEECO 930 MBE chamber, monitoring the sample with reflection high energy electron diffraction (RHEED) at 12 keV. After desorbing the native oxide on an epiready semi-insulating (001) GaAs substrate, we grew 30 nm of GaAs at a thermocouple temperature of 720 $\mathrm{^\circ C}$ using Ga (5N) and As (5N) evaporated from different uncracked effusion cells. The Ga (As) beam equivalent pressure was $5 \times 10^{-8} ~(7\times 10^{-7})$ torr. We then cooled down the substrate to a thermocouple temperature of 400 $\mathrm{^\circ C}$ in the presence of As flux. At this point, we observed a RHEED pattern showing a $2 \times 4$ GaAs surface reconstruction (Fig. 1 (a)). We then increased the temperature to 700-800 $\mathrm{^\circ C}$ and conjointly deposited As and Ta (the latter from a 4 pocket e-beam evaporator), obtaining the RHEED pattern shown in Fig. 1 (b) which displays characteristics of polycrystalline nature of the TaAs thin film (see Appendix A for details). We then transferred the TaAs films to an external electron beam evaporator for the deposition of \NiFe. Most of the samples studied here were transferred with a brief exposure to ambient, leading to an oxidized interface between TaAs and \NiFe. A few samples were transferred using a vacuum suitcase enabling the study of TaAs/\NiFe~heterostructures with a pristine interface. The complexity of the transfer procedure currently prevents us from doing a systematic study of unoxidized TaAs/\NiFe~heterostructures as a function of \NiFe~or TaAs layer thickness. The thickness of the \NiFe~was controlled by a quartz crystal monitor during growth, and the deposition rate of each layer was then confirmed {\it ex situ} using X-ray reflectivity measurements. We also grew simple metal Ta/\NiFe~heterostructures with and without natural oxidation on $\mathrm{SiO_2}$ substrates using electron beam evaporation for further spin torque analysis. All samples used in ST-FMR experiments were capped with 3 nm of Al to prevent oxidation of the \NiFe~layer. Our past studies show self-limited oxidation of such Al layers \cite{PhysRevApplied.16.054031yanez}. 

We studied the structural quality and crystalline phase of the TaAs films using X-ray diffraction. Figure 1 (c) shows a $\mathrm{2\theta-\omega}$ scan of a TaAs film, revealing (004), (008) and (112) peaks corresponding to the body centered tetragonal phase of TaAs. Measurements of the rocking curve around the (004) peak show a full width half maximum angle of 0.73$\mathrm{^\circ }$. A reciprocal space map around the TaAs (004) peak displays isotropic broadening, consistent with the polycrystalline nature of the film (Fig. 1 (d)). Figure 1 (e) shows the surface morphology of the film measured using atomic force microscopy (AFM). We observe several grains with a size of up to 100 nm $\times $ 50 nm. Height profiles of the AFM data in the vertical direction (Fig. 1 (f)) show 1 nm height variations between different grains that match the lattice constant of TaAs along the c-axis \cite{XuPhysRevLett.116.096801,XuTaAsArpesScience}. We also used electrical measurements in 10 nm and 30 nm thick TaAs films using a 1 mm x 0.5 mm Hall bar configuration and determined a resistivity in the range $\mathrm{124\ \mu\Omega\cdot cm}$ to $\mathrm{318\ \mu\Omega\cdot  cm}$. These values are 2 to 5 times larger than the values reported in bulk crystals of TaAs \cite{PhysRevX.5.031023chiralanom}. Magnetoresistance measurements at low temperature ($T = 2$ K) show signs of weak localization  and Hall effect measurements yield carrier densities on the order of $10^{21}-10^{22}\ \mathrm{cm}^{-3}$ (see Appendix).

\begin{figure}
\includegraphics[width=120mm]{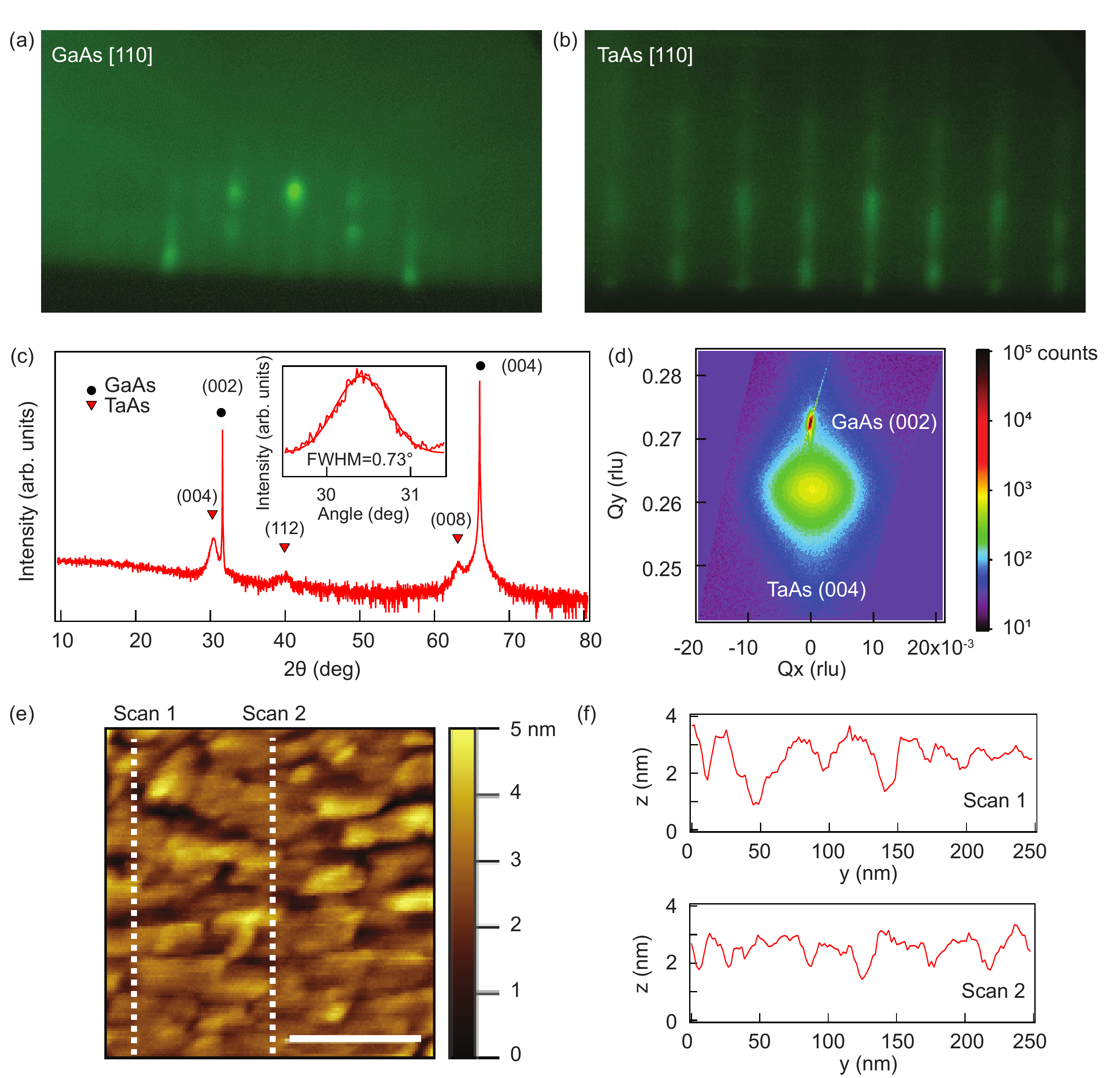}
\caption{\label{fig:1} RHEED pattern of the GaAs (a) and TaAs (b) layers along the [110] GaAs crystal direction. (c) X-ray diffraction $2\theta$ scan of a 30 nm thick TaAs film. The inset shows the rocking curve of the (004) TaAs peak, showing a full width half maximum of $0.73^\circ$. (d) Reciprocal space map of the TaAs thin film around the (004) peak in relative lattice units (rlu). (e) AFM image of the TaAs surface with a 100 nm scale bar. (f) Height profiles along the vertical directions shown in (e) of the TaAs film.}
\end{figure}

To obtain a deeper understanding of the microscopic structure of the TaAs thin films, we performed high-angle annular dark-field scanning transmission electron microscopy (HAADF-STEM). Figure 2(a) shows a lower magnification image of the film cross-section having neighboring crystalline regions of different orientations, corresponding to the grains of TaAs about 5-10 nm in size.  This can be better appreciated by taking the Fourier transform of smaller sections of the film showing unique crystal orientations, in order to identify the grain boundary regions as shown in Fig. 2(b). The polycrystalline nature of the TaAs film grown can possibly result from the large lattice mismatch between TaAs and GaAs, $17.5 \%$ in the $\langle 100 \rangle $ direction and $16.5 \%$ in the $\langle 110 \rangle$ direction. To obtain the chemical composition information from the film, we used energy dispersive X-ray (EDX) spectroscopy in the TEM measurements (Fig. 2(c)).  Elemental distribution across the heterostructure showed a 1:1 stoichiometry between Ta and As in the film region (Fig. 2(d)). Additionally, the 2 nm region of darker contrast seen in the HAADF images corresponds to the presence of an oxide layer of TaAsO$_x$, on the film surface also seen in the elemental maps. This is produced by the air exposure of the sample between the growth and its measurement (see Appendix F). This is further confirmed by X-ray spectroscopy from the surface of the film with take-off angles of $30^{\circ}$ and $80^{\circ}$ with respect to the sample surface plane. The spectra reveals the presence of Ta, As, Ta$_2$O$_5$ and AsO$_x$ on its surface (see Appendix C).

\begin{figure}
\includegraphics[width=160mm]{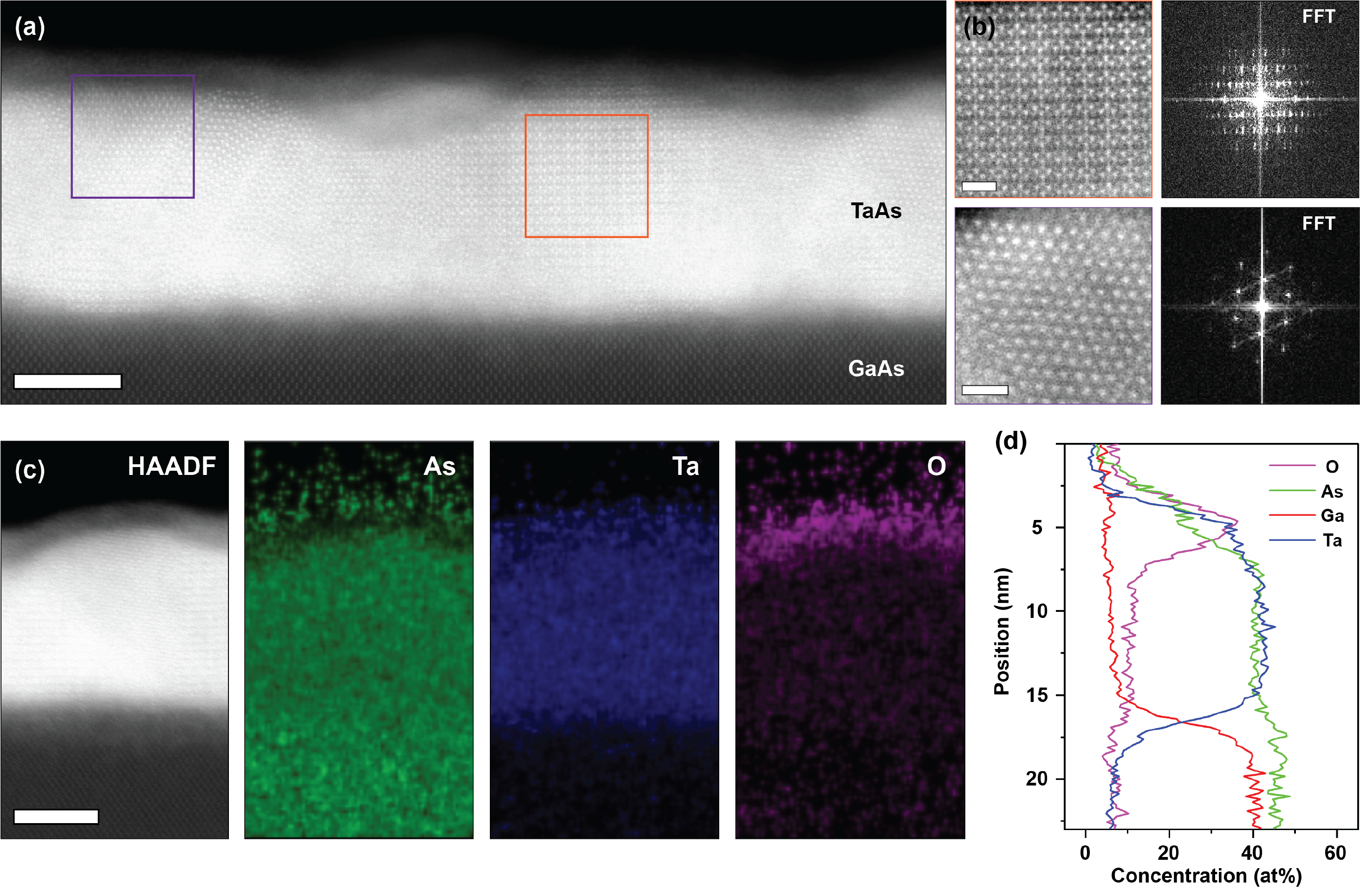}
\caption{\label{fig:2} (a) HAADF-STEM image of the heterostructure cross-section showing a $\sim 10$ nm polycrystalline TaAs film grown on GaAs layer. Scale bar is 5 nm. (b) Higher magnification STEM images of regions highlighted in a with their corresponding fast Fourier transforms (FFTs), showing the different grains present in the polycrystalline TaAs film layer highlighted in (a). Scale bars are 1 nm. (c) Elemental maps of As, Ta and O, showing the distribution of these elements across the film interface. Scale bar is 5 nm. (d) Concentration (atomic percent) of the Ta, Ga, As and O elements across the thin film section obtained from STEM-EDX.}

\end{figure}


We now discuss measurements of charge-spin conversion using ST-FMR, focusing on the naturally oxidized TaAs/\NiFe~heterostructures. We used standard lithography techniques to pattern the films into $\mathrm{50 ~\mu m \times 10 ~\mu m}$ bars and deposited Ti/Au contacts for electrical measurement (Fig. 3 (a)). We have measured a total of 16 devices including TaAs and Ta films with and without natural oxidation of its surface before the \NiFe~deposition. Amongst these, we include the data of one TaAs film with a pristine surface. To perform the ST-FMR experiments (Fig. 3 (b)), we apply a radio-frequency charge current ($J_c$) along the heterostructure, which produces a flow of angular momentum (spin current, $J_s$) in a direction perpendicular to the flow of charge via the spin Hall or Rashba-Edelstein effects. The spin accumulation in the TaAs/\NiFe~interface, after being absorbed into the ferromagnet, applies a torque on its magnetization causing it to precess. This phenomenon can then be measured as a rectified voltage across the device ($V_{\mathrm{mix}}$) due to the mixing of the applied current and the anisotropic magnetoresistance of the ferromagnet \cite{PhysRevLett.106.036601LiuSTFMR} (see Appendix D). This allows us to obtain the spectra shown in Figs. 3 (c) and 3 (d). In addition to the spin generated torques, the flow of charge also produces an Oersted field that interacts with the magnetization and produces an out-of-plane torque that can be used to quantify the spin torque efficiency \cite{PhysRevLett.106.036601LiuSTFMR,PhysRevB.92.064426CFPaiSTFMR}. 
We fit the resonance spectrum with a symmetric and antisymmetric Lorentzian contribution and determine the spin torque efficiency, which is given by \cite{PhysRevLett.106.036601LiuSTFMR,Mellnik2014,PhysRevApplied.16.054031yanez}: 
\begin{equation}
   \xi_{\mathrm{FMR}}= \frac{S}{A} \left(\frac{e}{\hbar}\right) \mu_0M_St_{\mathrm{TaAs}}t_{\mathrm{NiFe}} \left[1+\left(\frac{M_{\mathrm{Eff}}}{H_{\mathrm{Res}}}\right)\right]^{1/2}.
   \label{eq:xi}
\end{equation}

Here, $\xi_{\mathrm{FMR}}$ is the spin torque efficiency, $S$ ($A$) is the amplitude of the symmetric (antisymmetric) Lorentzian fit, $e$ is the charge of the electron, $\hbar$ is the reduced Planck constant, $\mu_0$ is the permeability of free space, $M_S=560\ \mathrm{kA/m}$ is the saturation magnetization of \NiFe~measured in a superconducting quantum interference device magnetometer), $t_\mathrm{{TaAs}}$ is the thickness of the TaAs layer, $t_\mathrm{{NiFe}}$ is the thickness of the ferromagnet, $M_{\mathrm{Eff}}$ is the effective magnetization obtained from fitting the Kittel FMR equation to the spectra and $H_\mathrm{{Res}}$ is the magnitude of magnetic field at which the resonance happens.

If the symmetric signal is produced solely by spin torque (in-plane) while the antisymmetric signal is produced by a combination of spin torque, Oersted field and other current-induced effective field-like torques (out-of-plane), we can write Eq. 1 as \cite{PhysRevB.92.064426CFPaiSTFMR}: 

\begin{equation}
  \frac{1}{\xi_{\mathrm{FMR}}}=\frac{1}{\xi_\mathrm{{DL}}} \left(1+\frac{\hbar}{e}\frac{\xi_\mathrm{{FL}}}{\mu_0M_St_{\mathrm{TaAs}}t_{\mathrm{NiFe}}}\right).
   \label{eq:xiinv}
\end{equation}

Here, $\xi_{\mathrm{DL}(\mathrm{FL})}=(2e/ \hbar)(4\pi M_St_{\mathrm{NiFe}})(H_{\mathrm{OP}(\mathrm{IP})}/ J_c)$ is the efficiency of the in-plane, damping-like (out-of-plane, field-like) torques, defined as the normalized ratio of the field $H_{\mathrm{OP}(\mathrm{IP})}$ generated in the out-of-plane (in-plane) directions  to the charge current flowing in the device \cite{PhysRevB.92.064426CFPaiSTFMR}. In systems where the current-induced effective field-like torque is negligible, the spin torque efficiency is equal to the damping-like torque efficiency ($\xi_{\mathrm{FMR}}=\xi_{\mathrm{DL}}$).

\begin{figure}
\includegraphics[width=110mm]{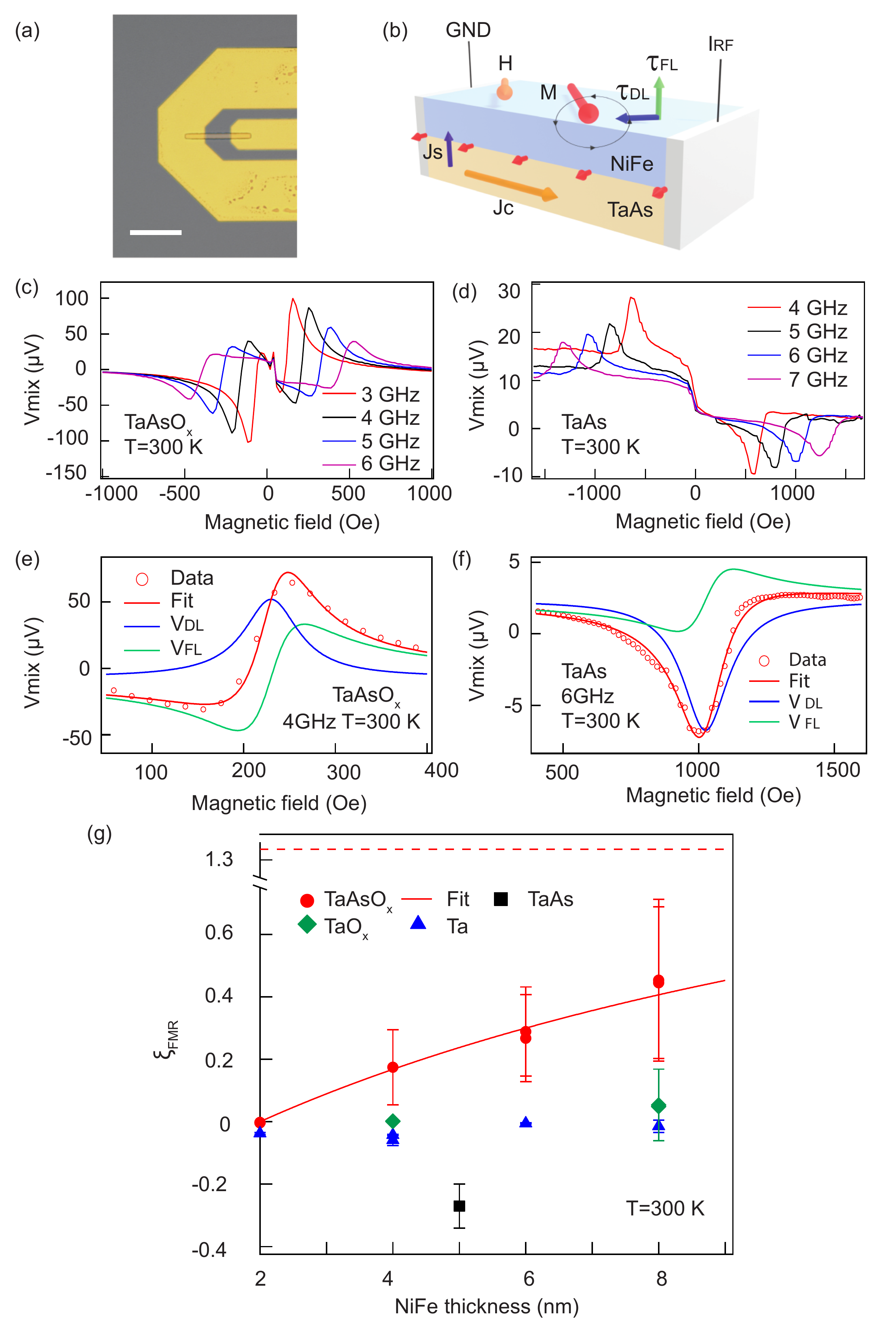}
\caption{\label{fig:3} (a) Optical image of a device used for ST-FMR measurements. The scale bar represents $\mathrm{100\ \mu m}$. (b) Schematic of the charge to spin conversion process in TaAs/\NiFe~heterostructures. Room temperature ST-FMR spectra measured at different excitation frequencies of (c) a 10 nm naturally oxidized TaAs polycrystalline film with 8 nm of \NiFe~and (d) and a non oxidized 10 nm TaAs polycrystalline film with 5 nm of \NiFe~(d). Fitted data of (e) the oxidized and (f) non oxidized samples described in (c) and (d), respectively. These show the damping-like symmetric ($V_\mathrm{DL}$) and field-like antisymmetric ($V_\mathrm{FL}$) Lorentzian contributions used in the fit. (g) Spin torque efficiency of TaAs and Ta with and without natural oxidation as a function of \NiFe~thickness. The solid line shows the fit of the $\mathrm{TaAsO_x}$ data and the dashed line its saturation value. }
\end{figure}

First, we focus on the naturally oxidized samples (abbreviated as $\mathrm{TaAsO_x}$ in the rest of this work). In Fig. 3 (c), we observe the spectra of a 10 nm oxidized $\mathrm{TaAs}$ film with an 8 nm \NiFe~layer. When we fit the data, we see a large symmetric component in the spectrum ($\mathrm{V_{DL}}$) that is larger than the antisymmetric one ($\mathrm{V_{FL}}$) and has the same sign as platinum control samples (not shown). We extract the spin torque efficiency of this heterostructure using Eq. 1 and obtain $\xi_\mathrm{{FMR}}=0.45\pm 0.25$. For further insight into the origin of this large spin torque efficiency, we grew a set of samples with fixed TaAs thickness (10 nm) and different \NiFe~thickness. We compute the spin torque efficiency of each sample and observe that it monotonically increases with the thickness of the ferromagnet (Fig. 3 (g)). This indicates the presence of a significant current-induced positive field-like torque \cite{YOuPhysRevB.94.140414} and a giant damping-like torque with an efficiency higher than our largest experimental value (see Appendix E). To quantify this phenomenon, we perform a linear fit of the inverse spin torque efficiency as a function of the inverse ferromagnet thickness considering a 2 nm dead layer in the ferromagnet and use Eq. 2 to determine the damping-like and field-like torque efficiencies. These values can be as large as $\xi_\mathrm{{DL}}=1.36\pm 0.66$ and $\xi_\mathrm{{FL}}=0.30\pm 0.15$ (see Appendix E). At present, we cannot determine the spin Hall conductivity of oxidized TaAs due to the difficulty of precisely determining the electrical conductivity of $\mathrm{TaAsO_x}$. Nevertheless, the value of the damping-like and field-like torque efficiency is comparable to that measured in topological insulators ($\mathrm{Bi_2Se_3}$) \cite{doi:10.1063/5.0035768HyunsooBi2se3,Mellnik2014,PhysRevLett.117.076601hailong} and higher than heavy metals (Pt, Ta)\cite{PhysRevB.92.064426CFPaiSTFMR,doi:10.1021/acsami.0c16485IrOx,doi:10.1021/acsami.1c00608HMOx}, Weyl semimetals ($\mathrm{WTe_2}$) \cite{doi:10.1063/1.5124688YiHongWTe2}, Dirac semimetals ($\mathrm{Cd_3As_2}$)\cite{PhysRevApplied.16.054031yanez} and Dirac nodal line semimetals ($\mathrm{IrO_2}$) \cite{doi:10.1021/acsami.0c16485IrOx}.

Next we focus on the pristine (unoxidized) TaAs/\NiFe~heterostructure. Figure 3 (f) shows that the symmetric component of the spectra changes sign with respect to the one taken in oxidized samples (Fig. 3 (e)). We follow the same procedure described above and obtain the spin torque efficiency of non oxidized TaAs $\xi_\mathrm{{FMR}}=-0.27\pm 0.07$. The sign of the spin torque efficiency is opposite to naturally oxidized TaAs and Pt control samples (see Appendix D). Assuming the effective field-like torque is negligible in the pristine TaAs sample, this sets a lower bound on the spin Hall angle ($\theta_{SH}$) due to the less than ideal spin transparency of the \NiFe~interface ($\theta_{SH}\geq\xi_\mathrm{{FMR}}$) \cite{PhysRevB.92.064426CFPaiSTFMR}. The spin torque efficiency can then be combined with our electrical conductivity measurements to obtain the spin Hall conductivity of TaAs. We find that in TaAs the lowest value of electrical conductivity measured in a Hall bar configuration is ($\sigma_{xx}=3144\ S\cdot cm^{-1}$) which implies that $|\sigma_{SH}|=\frac{\hbar}{2e}|\theta_{SH}|\sigma_{xx} \gtrsim 424\pm110\ \frac{\hbar}{e} \frac{\mathrm{S}}{\mathrm{cm}}$. This experimental value agrees with the theoretical prediction of a negative giant spin Hall conductivity in TaAs and quantitatively agrees with its value computed from first principles \cite{PhysRevLett.117.146403SHCtheoryBY}.

To study the effect of the oxide on the giant spin torque efficiency measured in TaAs films, we measured the ST-FMR spectra of simple metal Ta/\NiFe~heterostructures evaporated on $\mathrm{SiO_2}$ with and without natural oxidation. We kept the Ta layer fixed at 6 nm, varied the \NiFe~thickness, and computed the spin torque efficiency of each device (Fig. 3 (f)). Even though there was variation among samples due to the lack of precise control over the oxide layer, we find that oxidation st the interface changes the sign of the spin torque efficiency: it is positive in naturally oxidized Ta ($\mathrm{TaO_x/Ni_{0.8}Fe_{0.2}}$) ($\xi_\mathrm{{FMR}}=0.037\pm 0.030$) while it is negative in non-oxidized Ta/\NiFe~heterostrucures ($\xi_\mathrm{{FMR}}=-0.032\pm 0.009$). These values are one order of magnitude smaller than the ones measured in TaAs; thus, the oxide itself cannot alone explain the large spin torque efficiency measured in TaAs. Nevertheless, this comparison with control Ta layers show once again our qualitative finding that the presence of oxide can change the sign of the symmetric component in the ST-FMR spectrum.

In conclusion, we demonstrated that it is possible to deposit thin films of TaAs on GaAs in a way that is compatible with standard III-V semiconductor processing. This allows us to experimentally measure a giant spin torque efficiency in naturally oxidized and pristine polycrystalline TaAs/\NiFe~heterostructures. We observe the presence of large damping-like and field-like torques with efficiencies that are greater in magnitude than other materials of contemporary interest. \cite{doi:10.1063/5.0035768HyunsooBi2se3,Mellnik2014,PhysRevLett.117.076601hailong,PhysRevB.92.064426CFPaiSTFMR,doi:10.1021/acsami.0c16485IrOx,doi:10.1021/acsami.1c00608HMOx,doi:10.1063/1.5124688YiHongWTe2,PhysRevApplied.16.054031yanez,doi:10.1021/acsami.0c16485IrOx}. The extracted spin Hall conductivity of pristine TaAs is in remarkable agreement with theoretical predictions \cite{PhysRevLett.117.146403SHCtheoryBY}.  We show that the presence of an oxide produces a positive damping-like torque efficiency that changes the sign of the symmetric component of the ST-FMR spectrum. Several mechanisms have been identified as possible ways to produce spin torques due to oxidation, including enhanced spin Hall and Rashba-Edelstein effects \cite{An2016CuOx,PhysRevApplied.16.054031yanez}, material-dependent electron distribution near interfaces \cite{Tsai2018BiOx}, and an oxide-enhanced giant orbital Hall effect in transition metals \cite{PhysRevB.77.165117orbitalhall,PhysRevB.103.L121113OrbitalOxide}. The polycrystalline nature of our films may possibly also enhance the spin torque efficiency due to quantum confinement effects, akin to observations in topological insulators  \cite{DC2018}, but this is only speculation at present. The topological Weyl semimetal nature of TaAs with spin polarized surface states also likely plays a key role in enhancing the intrinsic spin Hall conductivity  \cite{PhysRevLett.117.146403SHCtheoryBY}. Determining and separating each of these contributions requires a higher control of the heterostructure that goes beyond the scope of this work. Nevertheless, our observation of a sign change in the ST-FMR spectra in oxidized samples is an easy and immediate way to discriminate between intrinsic effects and effects due to oxidation. The large spin torque efficiencies demonstrated in TaAs that can be further enhanced in the presence of oxide has immediate impact for fundamental research on spin-related phenomena in topological materials and in technological spintronic applications.

\begin{acknowledgments}

The authors would like to thank D. C. Ralph for valuable comments and A. Sengupta for providing access to apparatus used in ST-FMR measurements. The principal support for this project was provided by SMART, one of seven centers of nCORE, a Semiconductor Research Corporation program, sponsored by the National Institute of Standards and Technology (NIST). This supported the synthesis and standard characterization of TaAs heterostructures as well as charge-spin conversion measurements (WY, YO, JD, NS) and their characterization using STEM (SG, AM). Additional support for materials synthesis and characterization was provided by the Penn State Two-Dimensional Crystal Consortium-Materials Innovation Platform (2DCC-MIP) under NSF Grant No. DMR-2039351 (RX, ES, AR, NS). Parts of this work were carried out in the Characterization Facility, University of Minnesota, which receives partial support from the NSF through the MRSEC (Award Number DMR-2011401) and the NNCI (Award Number ECCS-2025124) programs.

\end{acknowledgments}

\newpage

\appendix

\section{\label{sec:level1}Growth of TaAs thin films and RHEED patterns along different crystal directions. }
TaAs crystallizes in a body-centered tetragonal structure ($\mathrm{I4_{1}md}$ space group) with a lattice constant of $a=0.3437$ nm and $c = 1.1656$ nm \cite{XuTaAsArpesScience,XuPhysRevLett.116.096801}. This makes it especially challenging to find an appropriate substrate for the thin film epitaxial growth of TaAs given that most commonly available semiconductors have lattice constants ranging from 0.541 nm  to 0.648 nm \cite{doi:10.1063/1.1368156bandgap}. We have tried to deposit TaAs in a variety of substrates including GaAs, MgO, $\mathrm{SrTiO_3}$, GaSb and $\mathrm{MgAl_2O_4}$ with the hope of stabilizing the film in a crystal orientation different than (001). Nevertheless, none of these samples show a clear RHEED pattern except for the films grown on GaAs; this suggests that GaAs has the correct surface chemistry to nucleate TaAs growth.  A second major obstacle in the synthesis of TaAs is the extremely high temperature ($\mathrm{\sim3000\ ^\circ C}$) required for Ta evaporation 
This cannot be achieved with a regular Knudsen effusion cell; if Ta is evaporated from a crucible in this temperature range, the vapor pressure of the crucible itself is significant enough to cause contamination in the sample. For these reasons, we evaporate Ta from 2 mm and 4 mm diameter rods using a 4-pocket electron beam evaporator. To achieve a reasonable Ta deposition rate, we increase the power applied to the rod until it melts and forms a controlled molten drop on one end. This allow us to achieve growth rates as high as 2.5 nm/h for the 2 mm rod and 10 nm/h for the 4 mm rod.  

\begin{figure}[b]
\includegraphics[width=120mm]{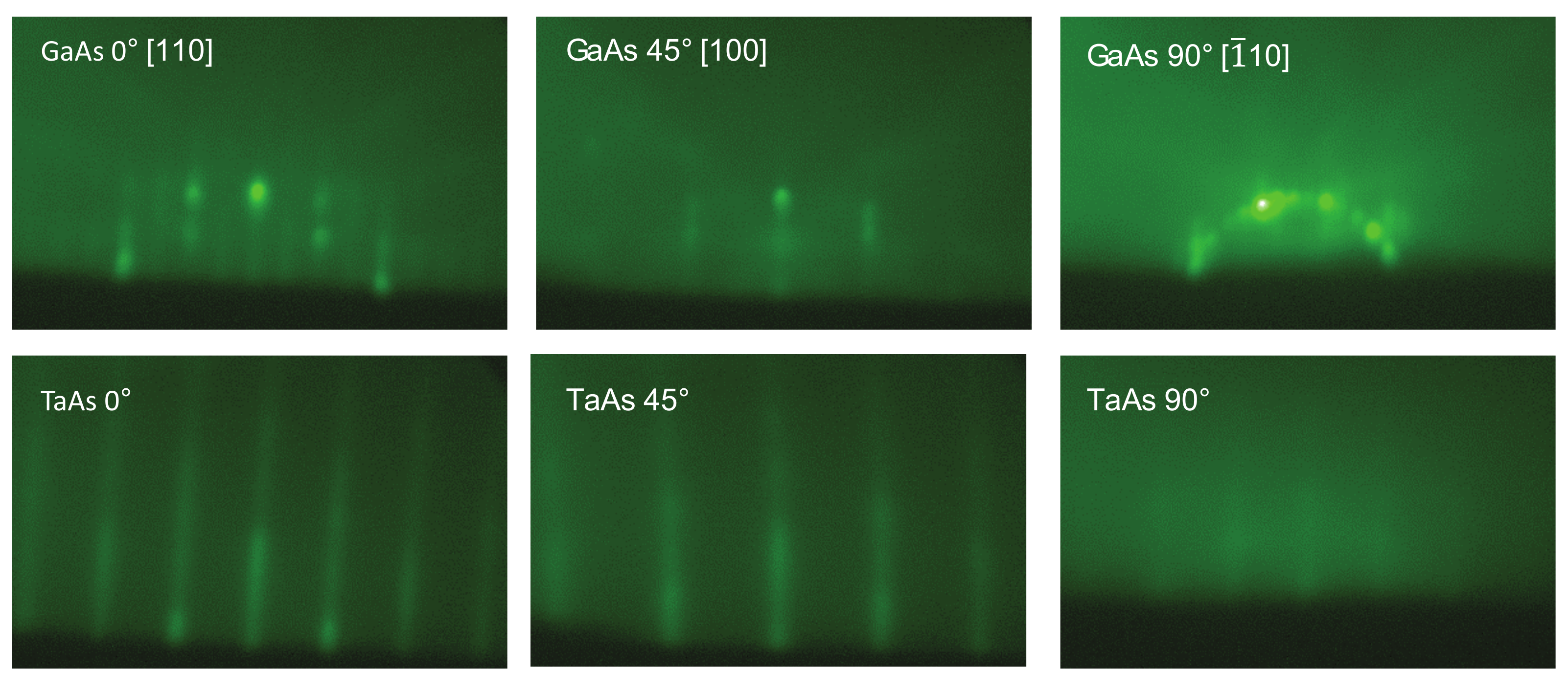}
\caption{\label{fig:s1} RHEED patterns of GaAs and TaAs along different crystal directions.}
\end{figure}

In Fig. 4, we show the RHEED patterns of GaAs and TaAs films along different crystal directions. GaAs shows a $2 \times 4$ surface reconstruction. In TaAs, the separation between fringes changes continuously as we rotate the sample, indicating polycrystalline growth. 

\section{Electrical characterization of TaAs in the Hall bar configuration.}

We used standard lithography techniques to pattern a 1 mm x 0.5 mm Hall bar (Fig. 5) and measured its resistance and Hall effect as a function of temperature and magnetic field. The resistivity of the sample monotonically increases with increasing temperature consistent with the expected metallic behavior of TaAs \cite{PhysRevX.5.031023chiralanom}. The residual resistivity ratio is small, consistent with the disorder in the polycrystalline film. At low temperature ($T \lesssim 10$K, the magnetoresistance shows signatures of weak localization. We computed the carrier density in the TaAs thin films using the Hall effect and obtained values on the order of $10^{21}-10^{22}\ \mathrm{cm}^{-3}$. 
\begin{figure}[ht]
\includegraphics[width=120mm]{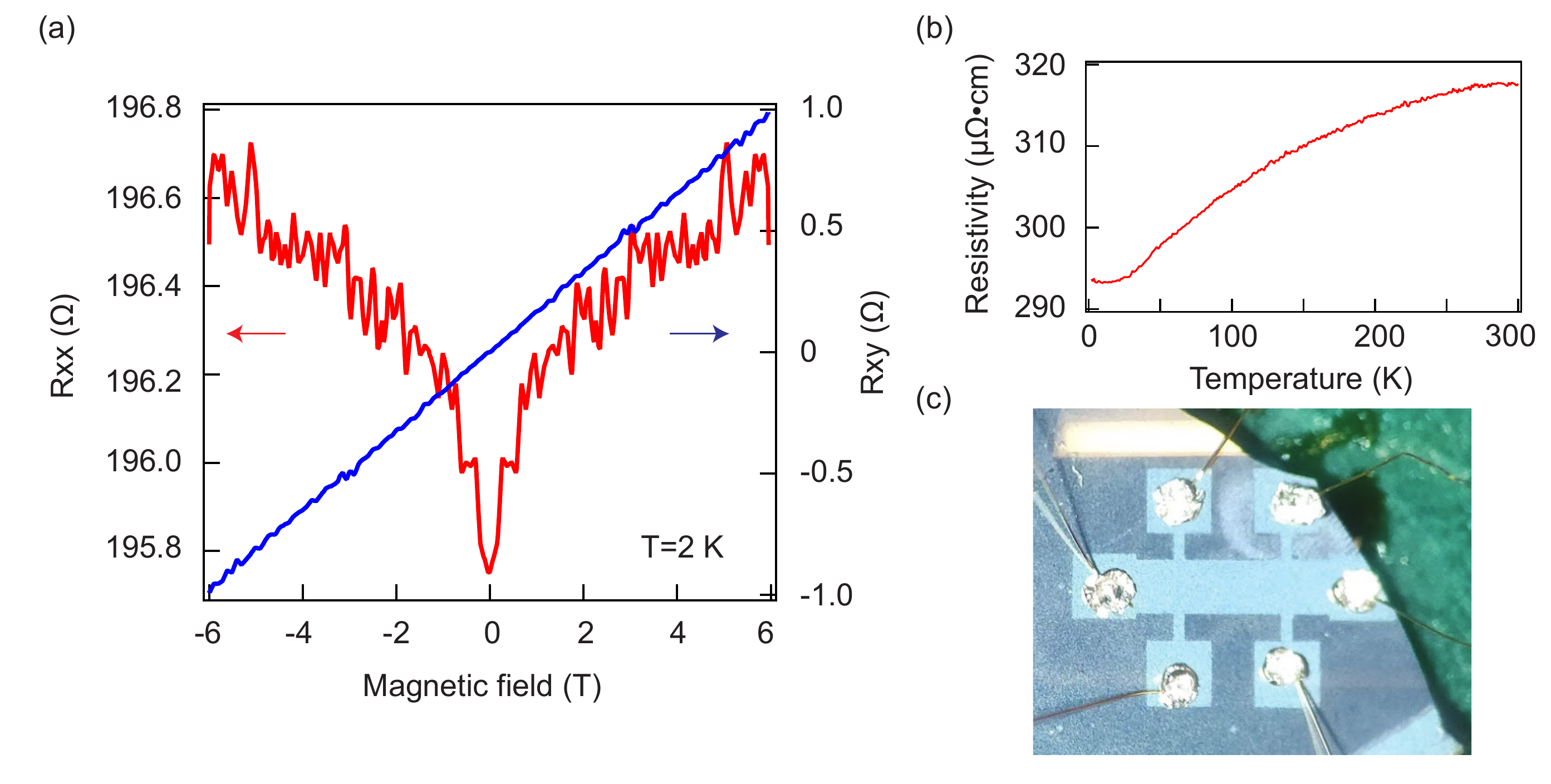}
\caption{\label{fig:s3} (a) Longitudinal ($R_{xx}$) and transverse ($R_{xy}$) resistance as a function of magnetic field at $T = 2$ K in a 30 nm thick TaAs film. (b) Resistivity of a 30 nm thick TaAs film as a function of temperature. (c) Optical image of the Hall bar used in these measurements.}
\end{figure}

\section{XPS data on TaAs thin films}
XPS experiments were performed on TaAs films of different thicknesses (Fig. 6) using a Physical Electronics VersaProbe II instrument equipped with a monochromatic Al $\mathrm{K\alpha}$ x-ray source ($\mathrm{h\nu = 1486.7\ eV}$) and a concentric hemispherical analyzer. Charge neutralization is performed using both low energy electrons ($\mathrm{<5 eV}$) and Ar ions. The binding energy axis is calibrated using sputter cleaned Cu (Cu $2p_{3/2} = 932.62$ eV, Cu $3p_{3/2} = 75.1$ eV) and Au foils (Au $4f_{7/2} = 83.96$ eV). Peaks were charge referenced to ${\mathrm{CH_x}}$ band in the carbon 1s spectra at 284.8 eV. Measurements are made at takeoff angles of 30° and 80° with respect to the sample surface plane. This results in a typical sampling depth of 2-3 nm and 4-6 nm respectively ($\mathrm{95\%}$ of the signal originated from this depth or shallower). Quantification is done using instrumental relative sensitivity factors that account for the x-ray cross section and inelastic mean free path of the electrons.

\begin{figure}[h]
\includegraphics[width=12cm]{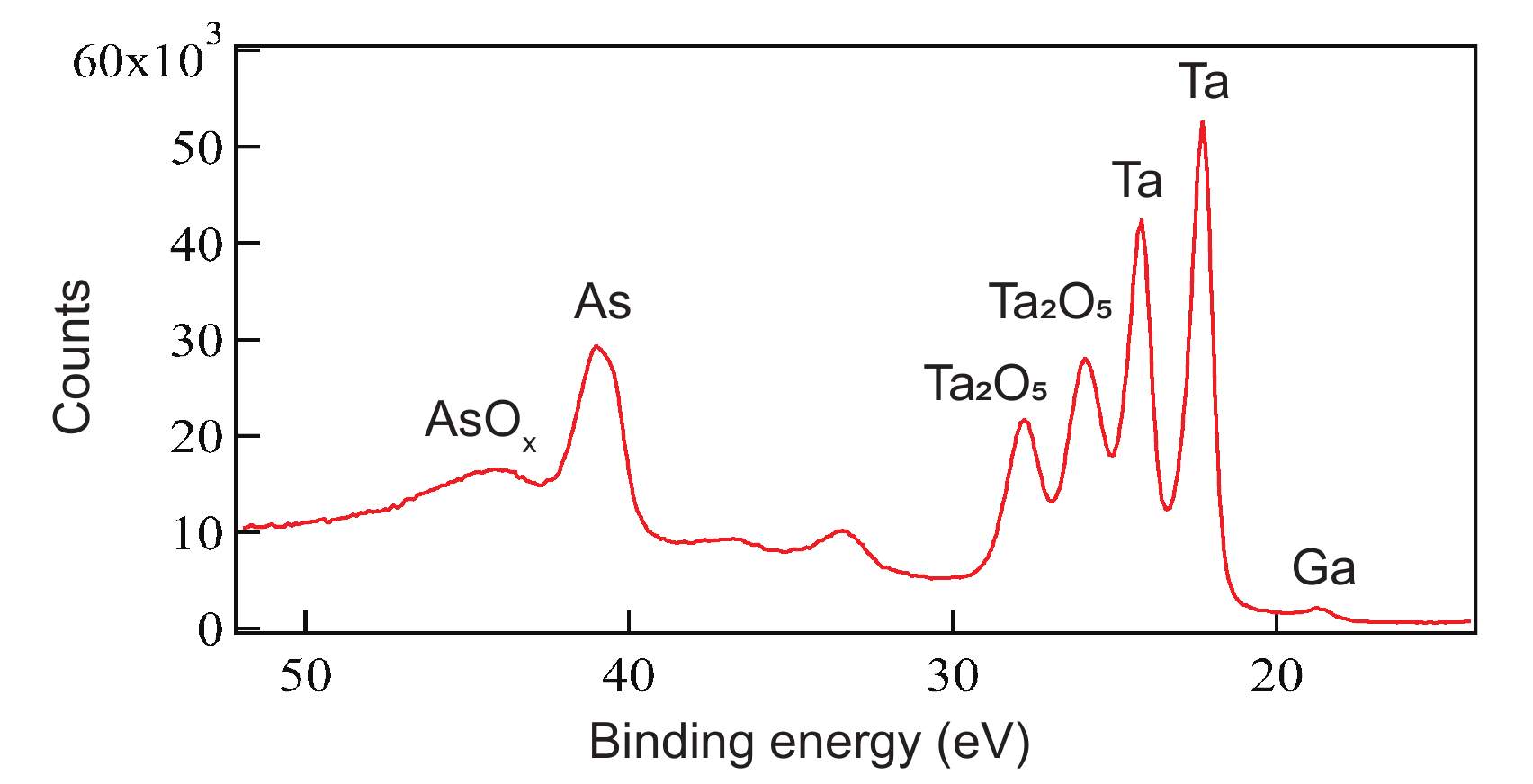}
\caption{\label{fig:s6} XPS data taken on a 10 nm thick oxidized TaAs film measured at 80 degrees with respect to the sample surface plane indicating the presence of $\mathrm{Ta_2O_5}$ and $\mathrm{AsO_x}$. }
\end{figure}

\section{ST-FMR measurements of oxidized and pristine Ta and TaAs.}

All the ST-FMR measurements are performed at room temperature using a Keithley 2182 nanovoltmeter, a radiofrequency current of 2 GHz to 7 GHz and an external magnetic field up to 1600 Oe. Figure 7 shows the ST-FMR spectrum of oxidized and pristine TaAs and Ta films that have been interfaced with \NiFe~films of different thickness. These signals are fitted using a symmetric and antisymmetric Lorentzian. The field-like antisymmetric contribution ($\mathrm{V_{FL}}$) has the same sign in all the samples while the symmetric contribution ($\mathrm{V_{DL}}$) changes sign in oxidized films in comparison with pristine ones. We also observe a large enhancement of spin torque efficiency in the pristine and oxidized TaAs films in comparison with pure Ta ones. This indicates that the spin torque signal measured in $\mathrm{TaAsO_x}$ cannot be explained by the presence of $\mathrm{TaO_x}$ alone.

\begin{figure}[ht]
\includegraphics[width=120mm]{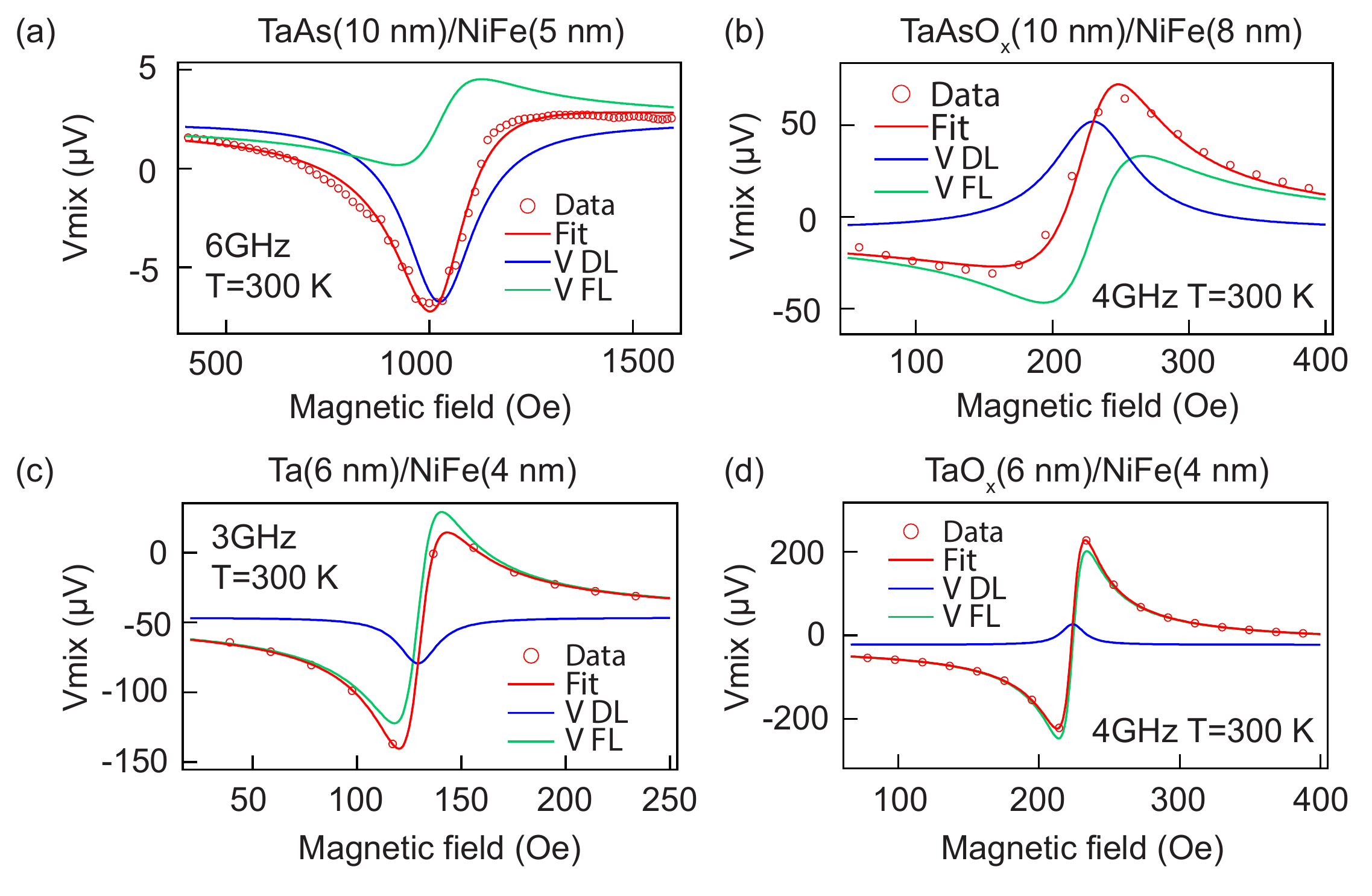}
\caption{\label{fig:s4} Room temperature ST-FMR spectrum of pristine (a) and oxidized TaAs (b) and pristine (c) and oxidized (d) Ta. The figure shows the fitted data showing the damping like symmetric($\mathrm{V_{DL}}$) and field like antisymmetric ($\mathrm{V_{FL}}$) Lorentzian contributions used in the fit.}
\end{figure}

\section{Analysis of the ST-FMR efficiency in oxidized TaAs samples with different ferromagnet thickness.}

To separate the contributions of the damping-like torque and other current-induced effective field-like torques in oxidized TaAs films \cite{PhysRevB.92.064426CFPaiSTFMR}, we have grown a set of samples where we kept the thickness of the TaAs layer constant at 10 nm and varied the thickness of the ferromagnetic \NiFe~layer. We notice that the spin torque efficiency is close to zero in samples with 2 nm of \NiFe. Together with a noticeable roughness in the TaAs surface (Fig. 2 (a) of the main text), this suggests the possible presence of a $\sim2$ nm dead magnetic layer at the TaAs/\NiFe~interface. To take this effect into account, we have subtracted the thickness of this dead layer from the \NiFe~thickness, as measured using a quartz crystal monitor during the deposition. This allows us to use Eq. E1 (Eq. 2 in the main text) to determine the damping-like and field-like torque efficiencies of oxidized TaAs. 
\begin{equation}
  \frac{1}{\xi_\mathrm{{FMR}}}=\frac{1}{\xi_\mathrm{{DL}}} \left(1+\frac{\hbar}{e}\frac{\xi_\mathrm{{FL}}}{\mu_0M_St_{\mathrm{{TaAs}}}t_{\mathrm{{NiFe}}}}\right).
   \label{eq:xiinvi}
\end{equation}
Here, $\xi_\mathrm{{FMR}}$ is the measured spin torque efficiency, $\xi_\mathrm{{DL(FL)}}=(2e/ \hbar)(J_{s, DL(FL)}/ J_c)$ is the spin torque efficiency of the in-plane, damping-like (out-of-plane, field-like) torques defined as the ratio of the spin current generating each torque by the amount of charge current flowing in the device, $e$ is the charge of the electron, $\hbar$ is the reduced Planck constant, $\mu_0$ is the permeability of free space, $M_S=560\ \mathrm{kA/m}$ is the saturation magnetization of \NiFe, $t_\mathrm{TaAs}$ is the thickness of the TaAs layer, $t_\mathrm{NiFe}$ is the effective thickness of the ferromagnet after subtracting the dead layer thickness (2 nm).

We perform a linear regression on the data shown in Fig. 8 (a) and obtained the efficiencies to be $\xi_\mathrm{{DL}}=1.36\pm 0.66$ and $\xi_\mathrm{{FL}}=0.30\pm 0.15$. This indicates the presence of a significant positive current-induced field-like torque on this system and a giant damping-like torque with an efficiency higher than our largest experimental value ($\xi_\mathrm{{FMR}}=0.45$ with 8 nm \NiFe. This can be seen better in Fig. 8 (b), in which we have used the parameters obtained from the linear fit to draw the nonlinear behavior of the spin torque efficiency as a function of ferromagnet thickness. We have performed a similar analysis ignoring the presence of the ferromagnetic dead layer and obtained an unphysical negative damping-like torque efficiency that does not agree with the positive damping-like torque (symmetric component of the ST-FMR spectra) measured in oxidized samples (Fig. 7 (a)).

\begin{figure}[ht]
\includegraphics[width=120mm]{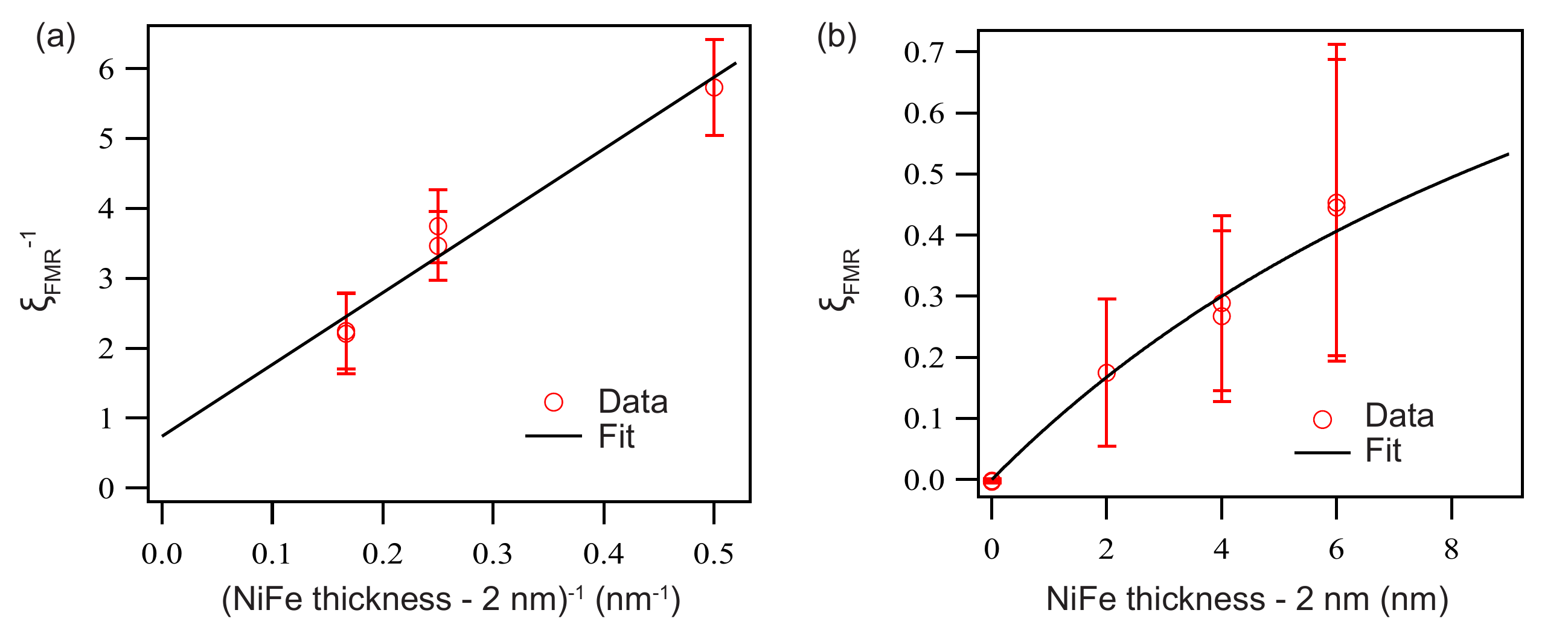}
\caption{\label{fig:s5} (a) Inverse spin torque efficiency as a function of inverse \NiFe~thickness for oxidized TaAs films showing the data and its linear fit. (b) Spin torque efficiency as a function of \NiFe~thickness for the same sample as (a). The values for the fit were determined from (a). In both panels, we subtracted 2 nm to the \NiFe~thickness to account for a possible ferromagnetic dead layer at the TaAs/\NiFe~interface.}
\end{figure}

\section{STEM Characterization of TaAs films} 
Samples for cross-sectional STEM characterization were prepared on a FEI Helios Nanolab G4 dual-beam focused ion beam (FIB) instrument. The sample was coated with amorphous carbon prior to sectioning, in order to prevent film surface damage by the Ga ion beam. The sample was thinned first using a 30 keV Ga ion beam follow by 2 keV Ga ion beam to remove the damage surface layers. 
High angle annular dark-field (HAADF)-STEM imaging and STEM-EDX spectroscopy were performed on am aberration-corrected FEI Titan G2 60-300 (S)TEM microscope, which is equipped with a CEOS DCOR probe corrector, a monochromator and a super-X energy dispersive X-ray (EDX) spectrometer. The microscope was operated at 200 keV with a probe convergence angle of 25.5 mrad. The inner and outer collection angles used for the HAADF detector are 55 and 200 mrad respectively. EDX maps were collected and analyzed using the Bruker Espirit 1.9 software.

\newpage



\begin{thebibliography}{35}%
\makeatletter
\providecommand \@ifxundefined [1]{%
 \@ifx{#1\undefined}
}%
\providecommand \@ifnum [1]{%
 \ifnum #1\expandafter \@firstoftwo
 \else \expandafter \@secondoftwo
 \fi
}%
\providecommand \@ifx [1]{%
 \ifx #1\expandafter \@firstoftwo
 \else \expandafter \@secondoftwo
 \fi
}%
\providecommand \natexlab [1]{#1}%
\providecommand \enquote  [1]{``#1''}%
\providecommand \bibnamefont  [1]{#1}%
\providecommand \bibfnamefont [1]{#1}%
\providecommand \citenamefont [1]{#1}%
\providecommand \href@noop [0]{\@secondoftwo}%
\providecommand \href [0]{\begingroup \@sanitize@url \@href}%
\providecommand \@href[1]{\@@startlink{#1}\@@href}%
\providecommand \@@href[1]{\endgroup#1\@@endlink}%
\providecommand \@sanitize@url [0]{\catcode `\\12\catcode `\$12\catcode
  `\&12\catcode `\#12\catcode `\^12\catcode `\_12\catcode `\%12\relax}%
\providecommand \@@startlink[1]{}%
\providecommand \@@endlink[0]{}%
\providecommand \url  [0]{\begingroup\@sanitize@url \@url }%
\providecommand \@url [1]{\endgroup\@href {#1}{\urlprefix }}%
\providecommand \urlprefix  [0]{URL }%
\providecommand \Eprint [0]{\href }%
\providecommand \doibase [0]{https://doi.org/}%
\providecommand \selectlanguage [0]{\@gobble}%
\providecommand \bibinfo  [0]{\@secondoftwo}%
\providecommand \bibfield  [0]{\@secondoftwo}%
\providecommand \translation [1]{[#1]}%
\providecommand \BibitemOpen [0]{}%
\providecommand \bibitemStop [0]{}%
\providecommand \bibitemNoStop [0]{.\EOS\space}%
\providecommand \EOS [0]{\spacefactor3000\relax}%
\providecommand \BibitemShut  [1]{\csname bibitem#1\endcsname}%
\let\auto@bib@innerbib\@empty
\bibitem [{\citenamefont {Liu}\ \emph {et~al.}(2012)\citenamefont {Liu},
  \citenamefont {Pai}, \citenamefont {Li}, \citenamefont {Tseng}, \citenamefont
  {Ralph},\ and\ \citenamefont {Buhrman}}]{doi:10.1126/science.1218197GiantTa}%
  \BibitemOpen
  \bibfield  {author} {\bibinfo {author} {\bibfnamefont {L.}~\bibnamefont
  {Liu}}, \bibinfo {author} {\bibfnamefont {C.-F.}\ \bibnamefont {Pai}},
  \bibinfo {author} {\bibfnamefont {Y.}~\bibnamefont {Li}}, \bibinfo {author}
  {\bibfnamefont {H.~W.}\ \bibnamefont {Tseng}}, \bibinfo {author}
  {\bibfnamefont {D.~C.}\ \bibnamefont {Ralph}},\ and\ \bibinfo {author}
  {\bibfnamefont {R.~A.}\ \bibnamefont {Buhrman}},\ }\bibfield  {title}
  {\bibinfo {title} {Spin-torque switching with the giant spin {H}all effect of
  {T}antalum},\ }\href {https://doi.org/10.1126/science.1218197} {\bibfield
  {journal} {\bibinfo  {journal} {Science}\ }\textbf {\bibinfo {volume}
  {336}},\ \bibinfo {pages} {555} (\bibinfo {year} {2012})}\BibitemShut
  {NoStop}%
\bibitem [{\citenamefont {Liu}\ \emph {et~al.}(2011)\citenamefont {Liu},
  \citenamefont {Moriyama}, \citenamefont {Ralph},\ and\ \citenamefont
  {Buhrman}}]{PhysRevLett.106.036601LiuSTFMR}%
  \BibitemOpen
  \bibfield  {author} {\bibinfo {author} {\bibfnamefont {L.}~\bibnamefont
  {Liu}}, \bibinfo {author} {\bibfnamefont {T.}~\bibnamefont {Moriyama}},
  \bibinfo {author} {\bibfnamefont {D.~C.}\ \bibnamefont {Ralph}},\ and\
  \bibinfo {author} {\bibfnamefont {R.~A.}\ \bibnamefont {Buhrman}},\
  }\bibfield  {title} {\bibinfo {title} {Spin-torque ferromagnetic resonance
  induced by the spin {H}all effect},\ }\href
  {https://doi.org/10.1103/PhysRevLett.106.036601} {\bibfield  {journal}
  {\bibinfo  {journal} {Phys. Rev. Lett.}\ }\textbf {\bibinfo {volume} {106}},\
  \bibinfo {pages} {036601} (\bibinfo {year} {2011})}\BibitemShut {NoStop}%
\bibitem [{\citenamefont {Pai}\ \emph {et~al.}(2015)\citenamefont {Pai},
  \citenamefont {Ou}, \citenamefont {Vilela-Le\~ao}, \citenamefont {Ralph},\
  and\ \citenamefont {Buhrman}}]{PhysRevB.92.064426CFPaiSTFMR}%
  \BibitemOpen
  \bibfield  {author} {\bibinfo {author} {\bibfnamefont {C.-F.}\ \bibnamefont
  {Pai}}, \bibinfo {author} {\bibfnamefont {Y.}~\bibnamefont {Ou}}, \bibinfo
  {author} {\bibfnamefont {L.~H.}\ \bibnamefont {Vilela-Le\~ao}}, \bibinfo
  {author} {\bibfnamefont {D.~C.}\ \bibnamefont {Ralph}},\ and\ \bibinfo
  {author} {\bibfnamefont {R.~A.}\ \bibnamefont {Buhrman}},\ }\bibfield
  {title} {\bibinfo {title} {Dependence of the efficiency of spin hall torque
  on the transparency of $\mathrm{Pt}$/ferromagnetic layer interfaces},\ }\href
  {https://doi.org/10.1103/PhysRevB.92.064426} {\bibfield  {journal} {\bibinfo
  {journal} {Phys. Rev. B}\ }\textbf {\bibinfo {volume} {92}},\ \bibinfo
  {pages} {064426} (\bibinfo {year} {2015})}\BibitemShut {NoStop}%
\bibitem [{\citenamefont {Mellnik}\ \emph {et~al.}(2014)\citenamefont
  {Mellnik}, \citenamefont {Lee}, \citenamefont {Richardella}, \citenamefont
  {Grab}, \citenamefont {Mintun}, \citenamefont {Fischer}, \citenamefont
  {Vaezi}, \citenamefont {Manchon}, \citenamefont {Kim}, \citenamefont
  {Samarth},\ and\ \citenamefont {Ralph}}]{Mellnik2014}%
  \BibitemOpen
  \bibfield  {author} {\bibinfo {author} {\bibfnamefont {A.~R.}\ \bibnamefont
  {Mellnik}}, \bibinfo {author} {\bibfnamefont {J.~S.}\ \bibnamefont {Lee}},
  \bibinfo {author} {\bibfnamefont {A.}~\bibnamefont {Richardella}}, \bibinfo
  {author} {\bibfnamefont {J.~L.}\ \bibnamefont {Grab}}, \bibinfo {author}
  {\bibfnamefont {P.~J.}\ \bibnamefont {Mintun}}, \bibinfo {author}
  {\bibfnamefont {M.~H.}\ \bibnamefont {Fischer}}, \bibinfo {author}
  {\bibfnamefont {A.}~\bibnamefont {Vaezi}}, \bibinfo {author} {\bibfnamefont
  {A.}~\bibnamefont {Manchon}}, \bibinfo {author} {\bibfnamefont {E.-A.}\
  \bibnamefont {Kim}}, \bibinfo {author} {\bibfnamefont {N.}~\bibnamefont
  {Samarth}},\ and\ \bibinfo {author} {\bibfnamefont {D.~C.}\ \bibnamefont
  {Ralph}},\ }\bibfield  {title} {\bibinfo {title} {Spin-transfer torque
  generated by a topological insulator},\ }\href
  {https://doi.org/10.1038/nature13534} {\bibfield  {journal} {\bibinfo
  {journal} {Nature}\ }\textbf {\bibinfo {volume} {511}},\ \bibinfo {pages}
  {449} (\bibinfo {year} {2014})}\BibitemShut {NoStop}%
\bibitem [{\citenamefont {Wang}\ \emph {et~al.}(2016)\citenamefont {Wang},
  \citenamefont {Kally}, \citenamefont {Lee}, \citenamefont {Liu},
  \citenamefont {Chang}, \citenamefont {Hickey}, \citenamefont {Mkhoyan},
  \citenamefont {Wu}, \citenamefont {Richardella},\ and\ \citenamefont
  {Samarth}}]{PhysRevLett.117.076601hailong}%
  \BibitemOpen
  \bibfield  {author} {\bibinfo {author} {\bibfnamefont {H.}~\bibnamefont
  {Wang}}, \bibinfo {author} {\bibfnamefont {J.}~\bibnamefont {Kally}},
  \bibinfo {author} {\bibfnamefont {J.~S.}\ \bibnamefont {Lee}}, \bibinfo
  {author} {\bibfnamefont {T.}~\bibnamefont {Liu}}, \bibinfo {author}
  {\bibfnamefont {H.}~\bibnamefont {Chang}}, \bibinfo {author} {\bibfnamefont
  {D.~R.}\ \bibnamefont {Hickey}}, \bibinfo {author} {\bibfnamefont {K.~A.}\
  \bibnamefont {Mkhoyan}}, \bibinfo {author} {\bibfnamefont {M.}~\bibnamefont
  {Wu}}, \bibinfo {author} {\bibfnamefont {A.}~\bibnamefont {Richardella}},\
  and\ \bibinfo {author} {\bibfnamefont {N.}~\bibnamefont {Samarth}},\
  }\bibfield  {title} {\bibinfo {title} {Surface-state-dominated spin-charge
  current conversion in topological-insulator/ferromagnetic-insulator
  heterostructures},\ }\href {https://doi.org/10.1103/PhysRevLett.117.076601}
  {\bibfield  {journal} {\bibinfo  {journal} {Phys. Rev. Lett.}\ }\textbf
  {\bibinfo {volume} {117}},\ \bibinfo {pages} {076601} (\bibinfo {year}
  {2016})}\BibitemShut {NoStop}%
\bibitem [{\citenamefont {Wang}\ \emph {et~al.}(2019)\citenamefont {Wang},
  \citenamefont {Kally}, \citenamefont {\ifmmode~\mbox{\c{S}}\else
  \c{S}\fi{}ahin}, \citenamefont {Liu}, \citenamefont {Yanez}, \citenamefont
  {Kamp}, \citenamefont {Richardella}, \citenamefont {Wu}, \citenamefont
  {Flatt\'e},\ and\ \citenamefont {Samarth}}]{PhysRevResearch.1.012014}%
  \BibitemOpen
  \bibfield  {author} {\bibinfo {author} {\bibfnamefont {H.}~\bibnamefont
  {Wang}}, \bibinfo {author} {\bibfnamefont {J.}~\bibnamefont {Kally}},
  \bibinfo {author} {\bibfnamefont {C.}~\bibnamefont
  {\ifmmode~\mbox{\c{S}}\else \c{S}\fi{}ahin}}, \bibinfo {author}
  {\bibfnamefont {T.}~\bibnamefont {Liu}}, \bibinfo {author} {\bibfnamefont
  {W.}~\bibnamefont {Yanez}}, \bibinfo {author} {\bibfnamefont {E.~J.}\
  \bibnamefont {Kamp}}, \bibinfo {author} {\bibfnamefont {A.}~\bibnamefont
  {Richardella}}, \bibinfo {author} {\bibfnamefont {M.}~\bibnamefont {Wu}},
  \bibinfo {author} {\bibfnamefont {M.~E.}\ \bibnamefont {Flatt\'e}},\ and\
  \bibinfo {author} {\bibfnamefont {N.}~\bibnamefont {Samarth}},\ }\bibfield
  {title} {\bibinfo {title} {Fermi level dependent spin pumping from a magnetic
  insulator into a topological insulator},\ }\href
  {https://doi.org/10.1103/PhysRevResearch.1.012014} {\bibfield  {journal}
  {\bibinfo  {journal} {Phys. Rev. Research}\ }\textbf {\bibinfo {volume}
  {1}},\ \bibinfo {pages} {012014(R)} (\bibinfo {year} {2019})}\BibitemShut
  {NoStop}%
\bibitem [{\citenamefont {Wang}\ \emph {et~al.}(2017)\citenamefont {Wang},
  \citenamefont {Zhu}, \citenamefont {Wu}, \citenamefont {Yang}, \citenamefont
  {Yu}, \citenamefont {Ramaswamy}, \citenamefont {Mishra}, \citenamefont {Shi},
  \citenamefont {Elyasi}, \citenamefont {Teo}, \citenamefont {Wu},\ and\
  \citenamefont {Yang}}]{Wang2017TIswitch}%
  \BibitemOpen
  \bibfield  {author} {\bibinfo {author} {\bibfnamefont {Y.}~\bibnamefont
  {Wang}}, \bibinfo {author} {\bibfnamefont {D.}~\bibnamefont {Zhu}}, \bibinfo
  {author} {\bibfnamefont {Y.}~\bibnamefont {Wu}}, \bibinfo {author}
  {\bibfnamefont {Y.}~\bibnamefont {Yang}}, \bibinfo {author} {\bibfnamefont
  {J.}~\bibnamefont {Yu}}, \bibinfo {author} {\bibfnamefont {R.}~\bibnamefont
  {Ramaswamy}}, \bibinfo {author} {\bibfnamefont {R.}~\bibnamefont {Mishra}},
  \bibinfo {author} {\bibfnamefont {S.}~\bibnamefont {Shi}}, \bibinfo {author}
  {\bibfnamefont {M.}~\bibnamefont {Elyasi}}, \bibinfo {author} {\bibfnamefont
  {K.-L.}\ \bibnamefont {Teo}}, \bibinfo {author} {\bibfnamefont
  {Y.}~\bibnamefont {Wu}},\ and\ \bibinfo {author} {\bibfnamefont
  {H.}~\bibnamefont {Yang}},\ }\bibfield  {title} {\bibinfo {title} {Room
  temperature magnetization switching in topological insulator-ferromagnet
  heterostructures by spin-orbit torques},\ }\href
  {https://doi.org/10.1038/s41467-017-01583-4} {\bibfield  {journal} {\bibinfo
  {journal} {Nat. Commun.}\ }\textbf {\bibinfo {volume} {8}},\ \bibinfo {pages}
  {1364} (\bibinfo {year} {2017})}\BibitemShut {NoStop}%
\bibitem [{\citenamefont {Wu}\ \emph {et~al.}(2019)\citenamefont {Wu},
  \citenamefont {Zhang}, \citenamefont {Deng}, \citenamefont {Lan},
  \citenamefont {Pan}, \citenamefont {Razavi}, \citenamefont {Che},
  \citenamefont {Huang}, \citenamefont {Dai}, \citenamefont {Wong},
  \citenamefont {Han},\ and\ \citenamefont {Wang}}]{PhysRevLett.123.207205}%
  \BibitemOpen
  \bibfield  {author} {\bibinfo {author} {\bibfnamefont {H.}~\bibnamefont
  {Wu}}, \bibinfo {author} {\bibfnamefont {P.}~\bibnamefont {Zhang}}, \bibinfo
  {author} {\bibfnamefont {P.}~\bibnamefont {Deng}}, \bibinfo {author}
  {\bibfnamefont {Q.}~\bibnamefont {Lan}}, \bibinfo {author} {\bibfnamefont
  {Q.}~\bibnamefont {Pan}}, \bibinfo {author} {\bibfnamefont {S.~A.}\
  \bibnamefont {Razavi}}, \bibinfo {author} {\bibfnamefont {X.}~\bibnamefont
  {Che}}, \bibinfo {author} {\bibfnamefont {L.}~\bibnamefont {Huang}}, \bibinfo
  {author} {\bibfnamefont {B.}~\bibnamefont {Dai}}, \bibinfo {author}
  {\bibfnamefont {K.}~\bibnamefont {Wong}}, \bibinfo {author} {\bibfnamefont
  {X.}~\bibnamefont {Han}},\ and\ \bibinfo {author} {\bibfnamefont {K.~L.}\
  \bibnamefont {Wang}},\ }\bibfield  {title} {\bibinfo {title}
  {Room-temperature spin-orbit torque from topological surface states},\ }\href
  {https://doi.org/10.1103/PhysRevLett.123.207205} {\bibfield  {journal}
  {\bibinfo  {journal} {Phys. Rev. Lett.}\ }\textbf {\bibinfo {volume} {123}},\
  \bibinfo {pages} {207205} (\bibinfo {year} {2019})}\BibitemShut {NoStop}%
\bibitem [{\citenamefont {Kondou}\ \emph {et~al.}(2016)\citenamefont {Kondou},
  \citenamefont {Yoshimi}, \citenamefont {Tsukazaki}, \citenamefont {Fukuma},
  \citenamefont {Matsuno}, \citenamefont {Takahashi}, \citenamefont {Kawasaki},
  \citenamefont {Tokura},\ and\ \citenamefont {Otani}}]{Kondou2016}%
  \BibitemOpen
  \bibfield  {author} {\bibinfo {author} {\bibfnamefont {K.}~\bibnamefont
  {Kondou}}, \bibinfo {author} {\bibfnamefont {R.}~\bibnamefont {Yoshimi}},
  \bibinfo {author} {\bibfnamefont {A.}~\bibnamefont {Tsukazaki}}, \bibinfo
  {author} {\bibfnamefont {Y.}~\bibnamefont {Fukuma}}, \bibinfo {author}
  {\bibfnamefont {J.}~\bibnamefont {Matsuno}}, \bibinfo {author} {\bibfnamefont
  {K.~S.}\ \bibnamefont {Takahashi}}, \bibinfo {author} {\bibfnamefont
  {M.}~\bibnamefont {Kawasaki}}, \bibinfo {author} {\bibfnamefont
  {Y.}~\bibnamefont {Tokura}},\ and\ \bibinfo {author} {\bibfnamefont
  {Y.}~\bibnamefont {Otani}},\ }\bibfield  {title} {\bibinfo {title}
  {Fermi-level-dependent charge-to-spin current conversion by {D}irac surface
  states of topological insulators},\ }\href
  {https://doi.org/10.1038/nphys3833} {\bibfield  {journal} {\bibinfo
  {journal} {Nat. Phys.}\ }\textbf {\bibinfo {volume} {12}},\ \bibinfo {pages}
  {1027} (\bibinfo {year} {2016})}\BibitemShut {NoStop}%
\bibitem [{\citenamefont {MacNeill}\ \emph
  {et~al.}(2017{\natexlab{a}})\citenamefont {MacNeill}, \citenamefont {Stiehl},
  \citenamefont {Guimaraes}, \citenamefont {Buhrman}, \citenamefont {Park},\
  and\ \citenamefont {Ralph}}]{MacNeill2017WTe2}%
  \BibitemOpen
  \bibfield  {author} {\bibinfo {author} {\bibfnamefont {D.}~\bibnamefont
  {MacNeill}}, \bibinfo {author} {\bibfnamefont {G.~M.}\ \bibnamefont
  {Stiehl}}, \bibinfo {author} {\bibfnamefont {M.~H.~D.}\ \bibnamefont
  {Guimaraes}}, \bibinfo {author} {\bibfnamefont {R.~A.}\ \bibnamefont
  {Buhrman}}, \bibinfo {author} {\bibfnamefont {J.}~\bibnamefont {Park}},\ and\
  \bibinfo {author} {\bibfnamefont {D.~C.}\ \bibnamefont {Ralph}},\ }\bibfield
  {title} {\bibinfo {title} {Control of spin--orbit torques through crystal
  symmetry in {WTe$_2$}/ferromagnet bilayers},\ }\href
  {https://doi.org/10.1038/nphys3933} {\bibfield  {journal} {\bibinfo
  {journal} {Nat. Phys.}\ }\textbf {\bibinfo {volume} {13}},\ \bibinfo {pages}
  {300} (\bibinfo {year} {2017}{\natexlab{a}})}\BibitemShut {NoStop}%
\bibitem [{\citenamefont {MacNeill}\ \emph
  {et~al.}(2017{\natexlab{b}})\citenamefont {MacNeill}, \citenamefont {Stiehl},
  \citenamefont {Guimar\~aes}, \citenamefont {Reynolds}, \citenamefont
  {Buhrman},\ and\ \citenamefont {Ralph}}]{PhysRevB.96.054450Wte2}%
  \BibitemOpen
  \bibfield  {author} {\bibinfo {author} {\bibfnamefont {D.}~\bibnamefont
  {MacNeill}}, \bibinfo {author} {\bibfnamefont {G.~M.}\ \bibnamefont
  {Stiehl}}, \bibinfo {author} {\bibfnamefont {M.~H.~D.}\ \bibnamefont
  {Guimar\~aes}}, \bibinfo {author} {\bibfnamefont {N.~D.}\ \bibnamefont
  {Reynolds}}, \bibinfo {author} {\bibfnamefont {R.~A.}\ \bibnamefont
  {Buhrman}},\ and\ \bibinfo {author} {\bibfnamefont {D.~C.}\ \bibnamefont
  {Ralph}},\ }\bibfield  {title} {\bibinfo {title} {Thickness dependence of
  spin-orbit torques generated by {WTe$_2$}},\ }\href
  {https://link.aps.org/doi/10.1103/PhysRevB.96.054450} {\bibfield  {journal}
  {\bibinfo  {journal} {Phys. Rev. B}\ }\textbf {\bibinfo {volume} {96}},\
  \bibinfo {pages} {054450} (\bibinfo {year} {2017}{\natexlab{b}})}\BibitemShut
  {NoStop}%
\bibitem [{\citenamefont {Shi}\ \emph {et~al.}(2019)\citenamefont {Shi},
  \citenamefont {Liang}, \citenamefont {Zhu}, \citenamefont {Cai},
  \citenamefont {Pollard}, \citenamefont {Wang}, \citenamefont {Wang},
  \citenamefont {Wang}, \citenamefont {He}, \citenamefont {Yu}, \citenamefont
  {Eda}, \citenamefont {Liang},\ and\ \citenamefont {Yang}}]{Shi2019}%
  \BibitemOpen
  \bibfield  {author} {\bibinfo {author} {\bibfnamefont {S.}~\bibnamefont
  {Shi}}, \bibinfo {author} {\bibfnamefont {S.}~\bibnamefont {Liang}}, \bibinfo
  {author} {\bibfnamefont {Z.}~\bibnamefont {Zhu}}, \bibinfo {author}
  {\bibfnamefont {K.}~\bibnamefont {Cai}}, \bibinfo {author} {\bibfnamefont
  {S.~D.}\ \bibnamefont {Pollard}}, \bibinfo {author} {\bibfnamefont
  {Y.}~\bibnamefont {Wang}}, \bibinfo {author} {\bibfnamefont {J.}~\bibnamefont
  {Wang}}, \bibinfo {author} {\bibfnamefont {Q.}~\bibnamefont {Wang}}, \bibinfo
  {author} {\bibfnamefont {P.}~\bibnamefont {He}}, \bibinfo {author}
  {\bibfnamefont {J.}~\bibnamefont {Yu}}, \bibinfo {author} {\bibfnamefont
  {G.}~\bibnamefont {Eda}}, \bibinfo {author} {\bibfnamefont {G.}~\bibnamefont
  {Liang}},\ and\ \bibinfo {author} {\bibfnamefont {H.}~\bibnamefont {Yang}},\
  }\bibfield  {title} {\bibinfo {title} {All-electric magnetization switching
  and {D}zyaloshinskii-{M}oriya interaction in $\mathrm{WTe_2}$/ferromagnet
  heterostructures},\ }\href {https://doi.org/10.1038/s41565-019-0525-8}
  {\bibfield  {journal} {\bibinfo  {journal} {Nat. Nano.}\ }\textbf {\bibinfo
  {volume} {14}},\ \bibinfo {pages} {945} (\bibinfo {year} {2019})}\BibitemShut
  {NoStop}%
\bibitem [{\citenamefont {Yanez}\ \emph {et~al.}(2021)\citenamefont {Yanez},
  \citenamefont {Ou}, \citenamefont {Xiao}, \citenamefont {Koo}, \citenamefont
  {Held}, \citenamefont {Ghosh}, \citenamefont {Rable}, \citenamefont
  {Pillsbury}, \citenamefont {Delgado}, \citenamefont {Yang}, \citenamefont
  {Chamorro}, \citenamefont {Grutter}, \citenamefont {Quarterman},
  \citenamefont {Richardella}, \citenamefont {Sengupta}, \citenamefont
  {McQueen}, \citenamefont {Borchers}, \citenamefont {Mkhoyan}, \citenamefont
  {Yan},\ and\ \citenamefont {Samarth}}]{PhysRevApplied.16.054031yanez}%
  \BibitemOpen
  \bibfield  {author} {\bibinfo {author} {\bibfnamefont {W.}~\bibnamefont
  {Yanez}}, \bibinfo {author} {\bibfnamefont {Y.}~\bibnamefont {Ou}}, \bibinfo
  {author} {\bibfnamefont {R.}~\bibnamefont {Xiao}}, \bibinfo {author}
  {\bibfnamefont {J.}~\bibnamefont {Koo}}, \bibinfo {author} {\bibfnamefont
  {J.~T.}\ \bibnamefont {Held}}, \bibinfo {author} {\bibfnamefont
  {S.}~\bibnamefont {Ghosh}}, \bibinfo {author} {\bibfnamefont
  {J.}~\bibnamefont {Rable}}, \bibinfo {author} {\bibfnamefont
  {T.}~\bibnamefont {Pillsbury}}, \bibinfo {author} {\bibfnamefont {E.~G.}\
  \bibnamefont {Delgado}}, \bibinfo {author} {\bibfnamefont {K.}~\bibnamefont
  {Yang}}, \bibinfo {author} {\bibfnamefont {J.}~\bibnamefont {Chamorro}},
  \bibinfo {author} {\bibfnamefont {A.~J.}\ \bibnamefont {Grutter}}, \bibinfo
  {author} {\bibfnamefont {P.}~\bibnamefont {Quarterman}}, \bibinfo {author}
  {\bibfnamefont {A.}~\bibnamefont {Richardella}}, \bibinfo {author}
  {\bibfnamefont {A.}~\bibnamefont {Sengupta}}, \bibinfo {author}
  {\bibfnamefont {T.}~\bibnamefont {McQueen}}, \bibinfo {author} {\bibfnamefont
  {J.~A.}\ \bibnamefont {Borchers}}, \bibinfo {author} {\bibfnamefont {K.~A.}\
  \bibnamefont {Mkhoyan}}, \bibinfo {author} {\bibfnamefont {B.}~\bibnamefont
  {Yan}},\ and\ \bibinfo {author} {\bibfnamefont {N.}~\bibnamefont {Samarth}},\
  }\bibfield  {title} {\bibinfo {title} {Spin and charge interconversion in
  {D}irac-semimetal thin films},\ }\href
  {https://link.aps.org/doi/10.1103/PhysRevApplied.16.054031} {\bibfield
  {journal} {\bibinfo  {journal} {Phys. Rev. Appl.}\ }\textbf {\bibinfo
  {volume} {16}},\ \bibinfo {pages} {054031} (\bibinfo {year}
  {2021})}\BibitemShut {NoStop}%
\bibitem [{\citenamefont {Xu}\ \emph {et~al.}(2020)\citenamefont {Xu},
  \citenamefont {Wei}, \citenamefont {Zhou}, \citenamefont {Feng},
  \citenamefont {Xu}, \citenamefont {Du}, \citenamefont {He}, \citenamefont
  {Huang}, \citenamefont {Zhang}, \citenamefont {Liu}, \citenamefont {Wu},
  \citenamefont {Guo}, \citenamefont {Wang}, \citenamefont {Guang},
  \citenamefont {Wei}, \citenamefont {Peng}, \citenamefont {Jiang},
  \citenamefont {Yu},\ and\ \citenamefont
  {Han}}]{https://doi.org/10.1002/adma.202000513xu}%
  \BibitemOpen
  \bibfield  {author} {\bibinfo {author} {\bibfnamefont {H.}~\bibnamefont
  {Xu}}, \bibinfo {author} {\bibfnamefont {J.}~\bibnamefont {Wei}}, \bibinfo
  {author} {\bibfnamefont {H.}~\bibnamefont {Zhou}}, \bibinfo {author}
  {\bibfnamefont {J.}~\bibnamefont {Feng}}, \bibinfo {author} {\bibfnamefont
  {T.}~\bibnamefont {Xu}}, \bibinfo {author} {\bibfnamefont {H.}~\bibnamefont
  {Du}}, \bibinfo {author} {\bibfnamefont {C.}~\bibnamefont {He}}, \bibinfo
  {author} {\bibfnamefont {Y.}~\bibnamefont {Huang}}, \bibinfo {author}
  {\bibfnamefont {J.}~\bibnamefont {Zhang}}, \bibinfo {author} {\bibfnamefont
  {Y.}~\bibnamefont {Liu}}, \bibinfo {author} {\bibfnamefont {H.-C.}\
  \bibnamefont {Wu}}, \bibinfo {author} {\bibfnamefont {C.}~\bibnamefont
  {Guo}}, \bibinfo {author} {\bibfnamefont {X.}~\bibnamefont {Wang}}, \bibinfo
  {author} {\bibfnamefont {Y.}~\bibnamefont {Guang}}, \bibinfo {author}
  {\bibfnamefont {H.}~\bibnamefont {Wei}}, \bibinfo {author} {\bibfnamefont
  {Y.}~\bibnamefont {Peng}}, \bibinfo {author} {\bibfnamefont {W.}~\bibnamefont
  {Jiang}}, \bibinfo {author} {\bibfnamefont {G.}~\bibnamefont {Yu}},\ and\
  \bibinfo {author} {\bibfnamefont {X.}~\bibnamefont {Han}},\ }\bibfield
  {title} {\bibinfo {title} {High spin hall conductivity in large-area
  type-{II} {D}irac semimetal $\mathrm{PtTe2}$},\ }\href
  {https://doi.org/https://doi.org/10.1002/adma.202000513} {\bibfield
  {journal} {\bibinfo  {journal} {Adv. Mater.}\ }\textbf {\bibinfo {volume}
  {32}},\ \bibinfo {pages} {2000513} (\bibinfo {year} {2020})}\BibitemShut
  {NoStop}%
\bibitem [{\citenamefont {Lin}\ \emph {et~al.}(2020)\citenamefont {Lin},
  \citenamefont {Wang}, \citenamefont {Wang}, \citenamefont {Li}, \citenamefont
  {Li}, \citenamefont {Xia}, \citenamefont {Yu},\ and\ \citenamefont
  {Liao}}]{PhysRevLett.124.116802FermiArcSpinCd3as2}%
  \BibitemOpen
  \bibfield  {author} {\bibinfo {author} {\bibfnamefont {B.-C.}\ \bibnamefont
  {Lin}}, \bibinfo {author} {\bibfnamefont {S.}~\bibnamefont {Wang}}, \bibinfo
  {author} {\bibfnamefont {A.-Q.}\ \bibnamefont {Wang}}, \bibinfo {author}
  {\bibfnamefont {Y.}~\bibnamefont {Li}}, \bibinfo {author} {\bibfnamefont
  {R.-R.}\ \bibnamefont {Li}}, \bibinfo {author} {\bibfnamefont
  {K.}~\bibnamefont {Xia}}, \bibinfo {author} {\bibfnamefont {D.}~\bibnamefont
  {Yu}},\ and\ \bibinfo {author} {\bibfnamefont {Z.-M.}\ \bibnamefont {Liao}},\
  }\bibfield  {title} {\bibinfo {title} {Electric control of {F}ermi arc spin
  transport in individual topological semimetal nanowires},\ }\href
  {https://doi.org/10.1103/PhysRevLett.124.116802} {\bibfield  {journal}
  {\bibinfo  {journal} {Phys. Rev. Lett.}\ }\textbf {\bibinfo {volume} {124}},\
  \bibinfo {pages} {116802} (\bibinfo {year} {2020})}\BibitemShut {NoStop}%
\bibitem [{\citenamefont {Stephen}\ \emph {et~al.}(2021)\citenamefont
  {Stephen}, \citenamefont {Hanbicki}, \citenamefont {Schumann}, \citenamefont
  {Robinson}, \citenamefont {Goyal}, \citenamefont {Stemmer},\ and\
  \citenamefont {Friedman}}]{doi:10.1021/acsnano.1c00154cd3as2spin}%
  \BibitemOpen
  \bibfield  {author} {\bibinfo {author} {\bibfnamefont {G.~M.}\ \bibnamefont
  {Stephen}}, \bibinfo {author} {\bibfnamefont {A.~T.}\ \bibnamefont
  {Hanbicki}}, \bibinfo {author} {\bibfnamefont {T.}~\bibnamefont {Schumann}},
  \bibinfo {author} {\bibfnamefont {J.~T.}\ \bibnamefont {Robinson}}, \bibinfo
  {author} {\bibfnamefont {M.}~\bibnamefont {Goyal}}, \bibinfo {author}
  {\bibfnamefont {S.}~\bibnamefont {Stemmer}},\ and\ \bibinfo {author}
  {\bibfnamefont {A.~L.}\ \bibnamefont {Friedman}},\ }\bibfield  {title}
  {\bibinfo {title} {Room-temperature spin transport in $\mathrm{Cd_3As_2}$},\
  }\href {https://doi.org/10.1021/acsnano.1c00154} {\bibfield  {journal}
  {\bibinfo  {journal} {ACS Nano}\ }\textbf {\bibinfo {volume} {15}},\ \bibinfo
  {pages} {5459} (\bibinfo {year} {2021})}\BibitemShut {NoStop}%
\bibitem [{\citenamefont {Xu}\ \emph {et~al.}(2015)\citenamefont {Xu},
  \citenamefont {Belopolski}, \citenamefont {Alidoust}, \citenamefont
  {Neupane}, \citenamefont {Bian}, \citenamefont {Zhang}, \citenamefont
  {Sankar}, \citenamefont {Chang}, \citenamefont {Yuan}, \citenamefont {Lee},
  \citenamefont {Huang}, \citenamefont {Zheng}, \citenamefont {Ma},
  \citenamefont {Sanchez}, \citenamefont {Wang}, \citenamefont {Bansil},
  \citenamefont {Chou}, \citenamefont {Shibayev}, \citenamefont {Lin},
  \citenamefont {Jia},\ and\ \citenamefont {Hasan}}]{XuTaAsArpesScience}%
  \BibitemOpen
  \bibfield  {author} {\bibinfo {author} {\bibfnamefont {S.-Y.}\ \bibnamefont
  {Xu}}, \bibinfo {author} {\bibfnamefont {I.}~\bibnamefont {Belopolski}},
  \bibinfo {author} {\bibfnamefont {N.}~\bibnamefont {Alidoust}}, \bibinfo
  {author} {\bibfnamefont {M.}~\bibnamefont {Neupane}}, \bibinfo {author}
  {\bibfnamefont {G.}~\bibnamefont {Bian}}, \bibinfo {author} {\bibfnamefont
  {C.}~\bibnamefont {Zhang}}, \bibinfo {author} {\bibfnamefont
  {R.}~\bibnamefont {Sankar}}, \bibinfo {author} {\bibfnamefont
  {G.}~\bibnamefont {Chang}}, \bibinfo {author} {\bibfnamefont
  {Z.}~\bibnamefont {Yuan}}, \bibinfo {author} {\bibfnamefont {C.-C.}\
  \bibnamefont {Lee}}, \bibinfo {author} {\bibfnamefont {S.-M.}\ \bibnamefont
  {Huang}}, \bibinfo {author} {\bibfnamefont {H.}~\bibnamefont {Zheng}},
  \bibinfo {author} {\bibfnamefont {J.}~\bibnamefont {Ma}}, \bibinfo {author}
  {\bibfnamefont {D.~S.}\ \bibnamefont {Sanchez}}, \bibinfo {author}
  {\bibfnamefont {B.}~\bibnamefont {Wang}}, \bibinfo {author} {\bibfnamefont
  {A.}~\bibnamefont {Bansil}}, \bibinfo {author} {\bibfnamefont
  {F.}~\bibnamefont {Chou}}, \bibinfo {author} {\bibfnamefont {P.~P.}\
  \bibnamefont {Shibayev}}, \bibinfo {author} {\bibfnamefont {H.}~\bibnamefont
  {Lin}}, \bibinfo {author} {\bibfnamefont {S.}~\bibnamefont {Jia}},\ and\
  \bibinfo {author} {\bibfnamefont {M.~Z.}\ \bibnamefont {Hasan}},\ }\bibfield
  {title} {\bibinfo {title} {Discovery of a {W}eyl fermion semimetal and
  topological {F}ermi arcs},\ }\href
  {https://doi.org/doi:10.1126/science.aaa9297TaAsArpes} {\bibfield  {journal}
  {\bibinfo  {journal} {Science}\ }\textbf {\bibinfo {volume} {349}},\ \bibinfo
  {pages} {613} (\bibinfo {year} {2015})}\BibitemShut {NoStop}%
\bibitem [{\citenamefont {Xu}\ \emph {et~al.}(2016)\citenamefont {Xu},
  \citenamefont {Belopolski}, \citenamefont {Sanchez}, \citenamefont {Neupane},
  \citenamefont {Chang}, \citenamefont {Yaji}, \citenamefont {Yuan},
  \citenamefont {Zhang}, \citenamefont {Kuroda}, \citenamefont {Bian},
  \citenamefont {Guo}, \citenamefont {Lu}, \citenamefont {Chang}, \citenamefont
  {Alidoust}, \citenamefont {Zheng}, \citenamefont {Lee}, \citenamefont
  {Huang}, \citenamefont {Hsu}, \citenamefont {Jeng}, \citenamefont {Bansil},
  \citenamefont {Neupert}, \citenamefont {Komori}, \citenamefont {Kondo},
  \citenamefont {Shin}, \citenamefont {Lin}, \citenamefont {Jia},\ and\
  \citenamefont {Hasan}}]{XuPhysRevLett.116.096801}%
  \BibitemOpen
  \bibfield  {author} {\bibinfo {author} {\bibfnamefont {S.-Y.}\ \bibnamefont
  {Xu}}, \bibinfo {author} {\bibfnamefont {I.}~\bibnamefont {Belopolski}},
  \bibinfo {author} {\bibfnamefont {D.~S.}\ \bibnamefont {Sanchez}}, \bibinfo
  {author} {\bibfnamefont {M.}~\bibnamefont {Neupane}}, \bibinfo {author}
  {\bibfnamefont {G.}~\bibnamefont {Chang}}, \bibinfo {author} {\bibfnamefont
  {K.}~\bibnamefont {Yaji}}, \bibinfo {author} {\bibfnamefont {Z.}~\bibnamefont
  {Yuan}}, \bibinfo {author} {\bibfnamefont {C.}~\bibnamefont {Zhang}},
  \bibinfo {author} {\bibfnamefont {K.}~\bibnamefont {Kuroda}}, \bibinfo
  {author} {\bibfnamefont {G.}~\bibnamefont {Bian}}, \bibinfo {author}
  {\bibfnamefont {C.}~\bibnamefont {Guo}}, \bibinfo {author} {\bibfnamefont
  {H.}~\bibnamefont {Lu}}, \bibinfo {author} {\bibfnamefont {T.-R.}\
  \bibnamefont {Chang}}, \bibinfo {author} {\bibfnamefont {N.}~\bibnamefont
  {Alidoust}}, \bibinfo {author} {\bibfnamefont {H.}~\bibnamefont {Zheng}},
  \bibinfo {author} {\bibfnamefont {C.-C.}\ \bibnamefont {Lee}}, \bibinfo
  {author} {\bibfnamefont {S.-M.}\ \bibnamefont {Huang}}, \bibinfo {author}
  {\bibfnamefont {C.-H.}\ \bibnamefont {Hsu}}, \bibinfo {author} {\bibfnamefont
  {H.-T.}\ \bibnamefont {Jeng}}, \bibinfo {author} {\bibfnamefont
  {A.}~\bibnamefont {Bansil}}, \bibinfo {author} {\bibfnamefont
  {T.}~\bibnamefont {Neupert}}, \bibinfo {author} {\bibfnamefont
  {F.}~\bibnamefont {Komori}}, \bibinfo {author} {\bibfnamefont
  {T.}~\bibnamefont {Kondo}}, \bibinfo {author} {\bibfnamefont
  {S.}~\bibnamefont {Shin}}, \bibinfo {author} {\bibfnamefont {H.}~\bibnamefont
  {Lin}}, \bibinfo {author} {\bibfnamefont {S.}~\bibnamefont {Jia}},\ and\
  \bibinfo {author} {\bibfnamefont {M.~Z.}\ \bibnamefont {Hasan}},\ }\bibfield
  {title} {\bibinfo {title} {Spin polarization and texture of the {F}ermi arcs
  in the {W}eyl fermion semimetal $\mathrm{TaAs}$},\ }\href
  {https://doi.org/10.1103/PhysRevLett.116.096801} {\bibfield  {journal}
  {\bibinfo  {journal} {Phys. Rev. Lett.}\ }\textbf {\bibinfo {volume} {116}},\
  \bibinfo {pages} {096801} (\bibinfo {year} {2016})}\BibitemShut {NoStop}%
\bibitem [{\citenamefont {Sun}\ \emph {et~al.}(2016)\citenamefont {Sun},
  \citenamefont {Zhang}, \citenamefont {Felser},\ and\ \citenamefont
  {Yan}}]{PhysRevLett.117.146403SHCtheoryBY}%
  \BibitemOpen
  \bibfield  {author} {\bibinfo {author} {\bibfnamefont {Y.}~\bibnamefont
  {Sun}}, \bibinfo {author} {\bibfnamefont {Y.}~\bibnamefont {Zhang}}, \bibinfo
  {author} {\bibfnamefont {C.}~\bibnamefont {Felser}},\ and\ \bibinfo {author}
  {\bibfnamefont {B.}~\bibnamefont {Yan}},\ }\bibfield  {title} {\bibinfo
  {title} {Strong intrinsic spin {Hall} effect in the {TaAs} family of {Weyl}
  semimetals},\ }\href {https://doi.org/10.1103/PhysRevLett.117.146403}
  {\bibfield  {journal} {\bibinfo  {journal} {Phys. Rev. Lett.}\ }\textbf
  {\bibinfo {volume} {117}},\ \bibinfo {pages} {146403} (\bibinfo {year}
  {2016})}\BibitemShut {NoStop}%
\bibitem [{\citenamefont {An}\ \emph {et~al.}(2016)\citenamefont {An},
  \citenamefont {Kageyama}, \citenamefont {Kanno}, \citenamefont {Enishi},\
  and\ \citenamefont {Ando}}]{An2016CuOx}%
  \BibitemOpen
  \bibfield  {author} {\bibinfo {author} {\bibfnamefont {H.}~\bibnamefont
  {An}}, \bibinfo {author} {\bibfnamefont {Y.}~\bibnamefont {Kageyama}},
  \bibinfo {author} {\bibfnamefont {Y.}~\bibnamefont {Kanno}}, \bibinfo
  {author} {\bibfnamefont {N.}~\bibnamefont {Enishi}},\ and\ \bibinfo {author}
  {\bibfnamefont {K.}~\bibnamefont {Ando}},\ }\bibfield  {title} {\bibinfo
  {title} {Spin--torque generator engineered by natural oxidation of
  $\mathrm{Cu}$},\ }\href {https://doi.org/10.1038/ncomms13069} {\bibfield
  {journal} {\bibinfo  {journal} {Nat. Commun.}\ }\textbf {\bibinfo {volume}
  {7}},\ \bibinfo {pages} {13069} (\bibinfo {year} {2016})}\BibitemShut
  {NoStop}%
\bibitem [{\citenamefont {Tsai}\ \emph {et~al.}(2018)\citenamefont {Tsai},
  \citenamefont {Karube}, \citenamefont {Kondou}, \citenamefont {Yamaguchi},
  \citenamefont {Ishii},\ and\ \citenamefont {Otani}}]{Tsai2018BiOx}%
  \BibitemOpen
  \bibfield  {author} {\bibinfo {author} {\bibfnamefont {H.}~\bibnamefont
  {Tsai}}, \bibinfo {author} {\bibfnamefont {S.}~\bibnamefont {Karube}},
  \bibinfo {author} {\bibfnamefont {K.}~\bibnamefont {Kondou}}, \bibinfo
  {author} {\bibfnamefont {N.}~\bibnamefont {Yamaguchi}}, \bibinfo {author}
  {\bibfnamefont {F.}~\bibnamefont {Ishii}},\ and\ \bibinfo {author}
  {\bibfnamefont {Y.}~\bibnamefont {Otani}},\ }\bibfield  {title} {\bibinfo
  {title} {Clear variation of spin splitting by changing electron distribution
  at non-magnetic metal/$\mathrm{Bi_2O_3}$ interfaces},\ }\href
  {https://doi.org/10.1038/s41598-018-23787-4} {\bibfield  {journal} {\bibinfo
  {journal} {Sci. Reports}\ }\textbf {\bibinfo {volume} {8}},\ \bibinfo {pages}
  {5564} (\bibinfo {year} {2018})}\BibitemShut {NoStop}%
\bibitem [{\citenamefont {Reyren}\ \emph {et~al.}(2012)\citenamefont {Reyren},
  \citenamefont {Bibes}, \citenamefont {Lesne}, \citenamefont {George},
  \citenamefont {Deranlot}, \citenamefont {Collin}, \citenamefont
  {Barth\'el\'emy},\ and\ \citenamefont
  {Jaffr\`es}}]{PhysRevLett.108.186802LAO1}%
  \BibitemOpen
  \bibfield  {author} {\bibinfo {author} {\bibfnamefont {N.}~\bibnamefont
  {Reyren}}, \bibinfo {author} {\bibfnamefont {M.}~\bibnamefont {Bibes}},
  \bibinfo {author} {\bibfnamefont {E.}~\bibnamefont {Lesne}}, \bibinfo
  {author} {\bibfnamefont {J.-M.}\ \bibnamefont {George}}, \bibinfo {author}
  {\bibfnamefont {C.}~\bibnamefont {Deranlot}}, \bibinfo {author}
  {\bibfnamefont {S.}~\bibnamefont {Collin}}, \bibinfo {author} {\bibfnamefont
  {A.}~\bibnamefont {Barth\'el\'emy}},\ and\ \bibinfo {author} {\bibfnamefont
  {H.}~\bibnamefont {Jaffr\`es}},\ }\bibfield  {title} {\bibinfo {title}
  {Gate-controlled spin injection at $\mathrm{LaAlO}_{3}/\mathrm{SrTiO}_{3}$
  interfaces},\ }\href {https://doi.org/10.1103/PhysRevLett.108.186802}
  {\bibfield  {journal} {\bibinfo  {journal} {Phys. Rev. Lett.}\ }\textbf
  {\bibinfo {volume} {108}},\ \bibinfo {pages} {186802} (\bibinfo {year}
  {2012})}\BibitemShut {NoStop}%
\bibitem [{\citenamefont {No{\"e}l}\ \emph {et~al.}(2020)\citenamefont
  {No{\"e}l}, \citenamefont {Trier}, \citenamefont {Vicente~Arche},
  \citenamefont {Br{\'e}hin}, \citenamefont {Vaz}, \citenamefont {Garcia},
  \citenamefont {Fusil}, \citenamefont {Barth{\'e}l{\'e}my}, \citenamefont
  {Vila}, \citenamefont {Bibes},\ and\ \citenamefont
  {Attan{\'e}}}]{Noel2020LAO2}%
  \BibitemOpen
  \bibfield  {author} {\bibinfo {author} {\bibfnamefont {P.}~\bibnamefont
  {No{\"e}l}}, \bibinfo {author} {\bibfnamefont {F.}~\bibnamefont {Trier}},
  \bibinfo {author} {\bibfnamefont {L.~M.}\ \bibnamefont {Vicente~Arche}},
  \bibinfo {author} {\bibfnamefont {J.}~\bibnamefont {Br{\'e}hin}}, \bibinfo
  {author} {\bibfnamefont {D.~C.}\ \bibnamefont {Vaz}}, \bibinfo {author}
  {\bibfnamefont {V.}~\bibnamefont {Garcia}}, \bibinfo {author} {\bibfnamefont
  {S.}~\bibnamefont {Fusil}}, \bibinfo {author} {\bibfnamefont
  {A.}~\bibnamefont {Barth{\'e}l{\'e}my}}, \bibinfo {author} {\bibfnamefont
  {L.}~\bibnamefont {Vila}}, \bibinfo {author} {\bibfnamefont {M.}~\bibnamefont
  {Bibes}},\ and\ \bibinfo {author} {\bibfnamefont {J.-P.}\ \bibnamefont
  {Attan{\'e}}},\ }\bibfield  {title} {\bibinfo {title} {Non-volatile electric
  control of spin--charge conversion in a $\mathrm{SrTiO_3}$ {R}ashba system},\
  }\href {https://doi.org/10.1038/s41586-020-2197-9} {\bibfield  {journal}
  {\bibinfo  {journal} {Nature}\ }\textbf {\bibinfo {volume} {580}},\ \bibinfo
  {pages} {483} (\bibinfo {year} {2020})}\BibitemShut {NoStop}%
\bibitem [{\citenamefont {Ohya}\ \emph {et~al.}(2020)\citenamefont {Ohya},
  \citenamefont {Araki}, \citenamefont {Anh}, \citenamefont {Kaneta},
  \citenamefont {Seki}, \citenamefont {Tabata},\ and\ \citenamefont
  {Tanaka}}]{PhysRevResearch.2.012014LaO3}%
  \BibitemOpen
  \bibfield  {author} {\bibinfo {author} {\bibfnamefont {S.}~\bibnamefont
  {Ohya}}, \bibinfo {author} {\bibfnamefont {D.}~\bibnamefont {Araki}},
  \bibinfo {author} {\bibfnamefont {L.~D.}\ \bibnamefont {Anh}}, \bibinfo
  {author} {\bibfnamefont {S.}~\bibnamefont {Kaneta}}, \bibinfo {author}
  {\bibfnamefont {M.}~\bibnamefont {Seki}}, \bibinfo {author} {\bibfnamefont
  {H.}~\bibnamefont {Tabata}},\ and\ \bibinfo {author} {\bibfnamefont
  {M.}~\bibnamefont {Tanaka}},\ }\bibfield  {title} {\bibinfo {title}
  {Efficient intrinsic spin-to-charge current conversion in an all-epitaxial
  single-crystal perovskite-oxide heterostructure of
  $\mathrm{L}{\mathrm{a}}_{0.67}\mathrm{S}{\mathrm{r}}_{0.33}\mathrm{Mn}{\mathrm{o}}_{3}/\mathrm{LaAl}{\mathrm{o}}_{3}/\mathrm{SrTi}{\mathrm{o}}_{3}$},\
  }\href {https://doi.org/10.1103/PhysRevResearch.2.012014} {\bibfield
  {journal} {\bibinfo  {journal} {Phys. Rev. Research}\ }\textbf {\bibinfo
  {volume} {2}},\ \bibinfo {pages} {012014(R)} (\bibinfo {year}
  {2020})}\BibitemShut {NoStop}%
\bibitem [{\citenamefont {Tanaka}\ \emph {et~al.}(2008)\citenamefont {Tanaka},
  \citenamefont {Kontani}, \citenamefont {Naito}, \citenamefont {Naito},
  \citenamefont {Hirashima}, \citenamefont {Yamada},\ and\ \citenamefont
  {Inoue}}]{PhysRevB.77.165117orbitalhall}%
  \BibitemOpen
  \bibfield  {author} {\bibinfo {author} {\bibfnamefont {T.}~\bibnamefont
  {Tanaka}}, \bibinfo {author} {\bibfnamefont {H.}~\bibnamefont {Kontani}},
  \bibinfo {author} {\bibfnamefont {M.}~\bibnamefont {Naito}}, \bibinfo
  {author} {\bibfnamefont {T.}~\bibnamefont {Naito}}, \bibinfo {author}
  {\bibfnamefont {D.~S.}\ \bibnamefont {Hirashima}}, \bibinfo {author}
  {\bibfnamefont {K.}~\bibnamefont {Yamada}},\ and\ \bibinfo {author}
  {\bibfnamefont {J.}~\bibnamefont {Inoue}},\ }\bibfield  {title} {\bibinfo
  {title} {Intrinsic spin {H}all effect and orbital {H}all effect in $4d$ and
  $5d$ transition metals},\ }\href {https://doi.org/10.1103/PhysRevB.77.165117}
  {\bibfield  {journal} {\bibinfo  {journal} {Phys. Rev. B}\ }\textbf {\bibinfo
  {volume} {77}},\ \bibinfo {pages} {165117} (\bibinfo {year}
  {2008})}\BibitemShut {NoStop}%
\bibitem [{\citenamefont {Go}\ \emph {et~al.}(2021)\citenamefont {Go},
  \citenamefont {Jo}, \citenamefont {Gao}, \citenamefont {Ando}, \citenamefont
  {Bl\"ugel}, \citenamefont {Lee},\ and\ \citenamefont
  {Mokrousov}}]{PhysRevB.103.L121113OrbitalOxide}%
  \BibitemOpen
  \bibfield  {author} {\bibinfo {author} {\bibfnamefont {D.}~\bibnamefont
  {Go}}, \bibinfo {author} {\bibfnamefont {D.}~\bibnamefont {Jo}}, \bibinfo
  {author} {\bibfnamefont {T.}~\bibnamefont {Gao}}, \bibinfo {author}
  {\bibfnamefont {K.}~\bibnamefont {Ando}}, \bibinfo {author} {\bibfnamefont
  {S.}~\bibnamefont {Bl\"ugel}}, \bibinfo {author} {\bibfnamefont {H.-W.}\
  \bibnamefont {Lee}},\ and\ \bibinfo {author} {\bibfnamefont {Y.}~\bibnamefont
  {Mokrousov}},\ }\bibfield  {title} {\bibinfo {title} {Orbital {R}ashba effect
  in a surface-oxidized $\mathrm{Cu}$ film},\ }\href
  {https://doi.org/10.1103/PhysRevB.103.L121113} {\bibfield  {journal}
  {\bibinfo  {journal} {Phys. Rev. B}\ }\textbf {\bibinfo {volume} {103}},\
  \bibinfo {pages} {L121113} (\bibinfo {year} {2021})}\BibitemShut {NoStop}%
\bibitem [{\citenamefont {Vurgaftman}\ \emph {et~al.}(2001)\citenamefont
  {Vurgaftman}, \citenamefont {Meyer},\ and\ \citenamefont
  {Ram-Mohan}}]{doi:10.1063/1.1368156bandgap}%
  \BibitemOpen
  \bibfield  {author} {\bibinfo {author} {\bibfnamefont {I.}~\bibnamefont
  {Vurgaftman}}, \bibinfo {author} {\bibfnamefont {J.~R.}\ \bibnamefont
  {Meyer}},\ and\ \bibinfo {author} {\bibfnamefont {L.~R.}\ \bibnamefont
  {Ram-Mohan}},\ }\bibfield  {title} {\bibinfo {title} {Band parameters for
  $\mathrm{III–V}$ compound semiconductors and their alloys},\ }\href
  {https://doi.org/10.1063/1.1368156} {\bibfield  {journal} {\bibinfo
  {journal} {J. Appl. Phys.}\ }\textbf {\bibinfo {volume} {89}},\ \bibinfo
  {pages} {5815} (\bibinfo {year} {2001})}\BibitemShut {NoStop}%
\bibitem [{\citenamefont {Bedoya-Pinto}\ \emph {et~al.}(2020)\citenamefont
  {Bedoya-Pinto}, \citenamefont {Pandeya}, \citenamefont {Liu}, \citenamefont
  {Deniz}, \citenamefont {Chang}, \citenamefont {Tan}, \citenamefont {Han},
  \citenamefont {Jena}, \citenamefont {Kostanovskiy},\ and\ \citenamefont
  {Parkin}}]{Bedoya-Pinto_ACSNano}%
  \BibitemOpen
  \bibfield  {author} {\bibinfo {author} {\bibfnamefont {A.}~\bibnamefont
  {Bedoya-Pinto}}, \bibinfo {author} {\bibfnamefont {A.~K.}\ \bibnamefont
  {Pandeya}}, \bibinfo {author} {\bibfnamefont {D.}~\bibnamefont {Liu}},
  \bibinfo {author} {\bibfnamefont {H.}~\bibnamefont {Deniz}}, \bibinfo
  {author} {\bibfnamefont {K.}~\bibnamefont {Chang}}, \bibinfo {author}
  {\bibfnamefont {H.}~\bibnamefont {Tan}}, \bibinfo {author} {\bibfnamefont
  {H.}~\bibnamefont {Han}}, \bibinfo {author} {\bibfnamefont {J.}~\bibnamefont
  {Jena}}, \bibinfo {author} {\bibfnamefont {I.}~\bibnamefont {Kostanovskiy}},\
  and\ \bibinfo {author} {\bibfnamefont {S.~S.~P.}\ \bibnamefont {Parkin}},\
  }\bibfield  {title} {\bibinfo {title} {Realization of epitaxial nbp and tap
  weyl semimetal thin films},\ }\href
  {https://pubs.acs.org/doi/abs/10.1021/acsami.0c16485} {\bibfield  {journal}
  {\bibinfo  {journal} {ACS Nano}\ }\textbf {\bibinfo {volume} {14}},\ \bibinfo
  {pages} {4405} (\bibinfo {year} {2020})}\BibitemShut {NoStop}%
\bibitem [{\citenamefont {Huang}\ \emph {et~al.}(2015)\citenamefont {Huang},
  \citenamefont {Zhao}, \citenamefont {Long}, \citenamefont {Wang},
  \citenamefont {Chen}, \citenamefont {Yang}, \citenamefont {Liang},
  \citenamefont {Xue}, \citenamefont {Weng}, \citenamefont {Fang},
  \citenamefont {Dai},\ and\ \citenamefont
  {Chen}}]{PhysRevX.5.031023chiralanom}%
  \BibitemOpen
  \bibfield  {author} {\bibinfo {author} {\bibfnamefont {X.}~\bibnamefont
  {Huang}}, \bibinfo {author} {\bibfnamefont {L.}~\bibnamefont {Zhao}},
  \bibinfo {author} {\bibfnamefont {Y.}~\bibnamefont {Long}}, \bibinfo {author}
  {\bibfnamefont {P.}~\bibnamefont {Wang}}, \bibinfo {author} {\bibfnamefont
  {D.}~\bibnamefont {Chen}}, \bibinfo {author} {\bibfnamefont {Z.}~\bibnamefont
  {Yang}}, \bibinfo {author} {\bibfnamefont {H.}~\bibnamefont {Liang}},
  \bibinfo {author} {\bibfnamefont {M.}~\bibnamefont {Xue}}, \bibinfo {author}
  {\bibfnamefont {H.}~\bibnamefont {Weng}}, \bibinfo {author} {\bibfnamefont
  {Z.}~\bibnamefont {Fang}}, \bibinfo {author} {\bibfnamefont {X.}~\bibnamefont
  {Dai}},\ and\ \bibinfo {author} {\bibfnamefont {G.}~\bibnamefont {Chen}},\
  }\bibfield  {title} {\bibinfo {title} {Observation of the
  chiral-anomaly-induced negative magnetoresistance in 3d {W}eyl semimetal
  $\mathrm{TaAs}$},\ }\href {https://doi.org/10.1103/PhysRevX.5.031023}
  {\bibfield  {journal} {\bibinfo  {journal} {Phys. Rev. X}\ }\textbf {\bibinfo
  {volume} {5}},\ \bibinfo {pages} {031023} (\bibinfo {year}
  {2015})}\BibitemShut {NoStop}%
\bibitem [{\citenamefont {Ou}\ \emph {et~al.}(2016)\citenamefont {Ou},
  \citenamefont {Pai}, \citenamefont {Shi}, \citenamefont {Ralph},\ and\
  \citenamefont {Buhrman}}]{YOuPhysRevB.94.140414}%
  \BibitemOpen
  \bibfield  {author} {\bibinfo {author} {\bibfnamefont {Y.}~\bibnamefont
  {Ou}}, \bibinfo {author} {\bibfnamefont {C.-F.}\ \bibnamefont {Pai}},
  \bibinfo {author} {\bibfnamefont {S.}~\bibnamefont {Shi}}, \bibinfo {author}
  {\bibfnamefont {D.~C.}\ \bibnamefont {Ralph}},\ and\ \bibinfo {author}
  {\bibfnamefont {R.~A.}\ \bibnamefont {Buhrman}},\ }\bibfield  {title}
  {\bibinfo {title} {Origin of field-like spin-orbit torques in heavy
  metal/ferromagnet/oxide thin film heterostructures},\ }\href
  {https://doi.org/10.1103/PhysRevB.94.140414} {\bibfield  {journal} {\bibinfo
  {journal} {Phys. Rev. B}\ }\textbf {\bibinfo {volume} {94}},\ \bibinfo
  {pages} {140414(R)} (\bibinfo {year} {2016})}\BibitemShut {NoStop}%
\bibitem [{\citenamefont {Zhu}\ \emph {et~al.}(2021)\citenamefont {Zhu},
  \citenamefont {Wang}, \citenamefont {Shi}, \citenamefont {Teo}, \citenamefont
  {Wu},\ and\ \citenamefont {Yang}}]{doi:10.1063/5.0035768HyunsooBi2se3}%
  \BibitemOpen
  \bibfield  {author} {\bibinfo {author} {\bibfnamefont {D.}~\bibnamefont
  {Zhu}}, \bibinfo {author} {\bibfnamefont {Y.}~\bibnamefont {Wang}}, \bibinfo
  {author} {\bibfnamefont {S.}~\bibnamefont {Shi}}, \bibinfo {author}
  {\bibfnamefont {K.-L.}\ \bibnamefont {Teo}}, \bibinfo {author} {\bibfnamefont
  {Y.}~\bibnamefont {Wu}},\ and\ \bibinfo {author} {\bibfnamefont
  {H.}~\bibnamefont {Yang}},\ }\bibfield  {title} {\bibinfo {title} {Highly
  efficient charge-to-spin conversion from {\it in situ} $\mathrm{Bi_2Se_3/Fe}$
  heterostructures},\ }\href {https://doi.org/10.1063/5.0035768} {\bibfield
  {journal} {\bibinfo  {journal} {Appl. Phys. Lett.}\ }\textbf {\bibinfo
  {volume} {118}},\ \bibinfo {pages} {062403} (\bibinfo {year}
  {2021})}\BibitemShut {NoStop}%
\bibitem [{\citenamefont {Bose}\ \emph {et~al.}(2020)\citenamefont {Bose},
  \citenamefont {Nelson}, \citenamefont {Zhang}, \citenamefont {Jadaun},
  \citenamefont {Jain}, \citenamefont {Schlom}, \citenamefont {Ralph},
  \citenamefont {Muller}, \citenamefont {Shen},\ and\ \citenamefont
  {Buhrman}}]{doi:10.1021/acsami.0c16485IrOx}%
  \BibitemOpen
  \bibfield  {author} {\bibinfo {author} {\bibfnamefont {A.}~\bibnamefont
  {Bose}}, \bibinfo {author} {\bibfnamefont {J.~N.}\ \bibnamefont {Nelson}},
  \bibinfo {author} {\bibfnamefont {X.~S.}\ \bibnamefont {Zhang}}, \bibinfo
  {author} {\bibfnamefont {P.}~\bibnamefont {Jadaun}}, \bibinfo {author}
  {\bibfnamefont {R.}~\bibnamefont {Jain}}, \bibinfo {author} {\bibfnamefont
  {D.~G.}\ \bibnamefont {Schlom}}, \bibinfo {author} {\bibfnamefont {D.~C.}\
  \bibnamefont {Ralph}}, \bibinfo {author} {\bibfnamefont {D.~A.}\ \bibnamefont
  {Muller}}, \bibinfo {author} {\bibfnamefont {K.~M.}\ \bibnamefont {Shen}},\
  and\ \bibinfo {author} {\bibfnamefont {R.~A.}\ \bibnamefont {Buhrman}},\
  }\bibfield  {title} {\bibinfo {title} {Effects of anisotropic strain on
  spin–orbit torque produced by the {D}irac nodal line semimetal
  $\mathrm{IrO_2}$},\ }\href
  {https://pubs.acs.org/doi/abs/10.1021/acsami.0c16485} {\bibfield  {journal}
  {\bibinfo  {journal} {ACS Appl. Mater. Interfaces}\ }\textbf {\bibinfo
  {volume} {12}},\ \bibinfo {pages} {55411} (\bibinfo {year}
  {2020})}\BibitemShut {NoStop}%
\bibitem [{\citenamefont {Lee}\ \emph {et~al.}(2021)\citenamefont {Lee},
  \citenamefont {Jeong}, \citenamefont {Yun}, \citenamefont {Park},
  \citenamefont {Ju}, \citenamefont {Lee}, \citenamefont {Min}, \citenamefont
  {Koo},\ and\ \citenamefont {Lee}}]{doi:10.1021/acsami.1c00608HMOx}%
  \BibitemOpen
  \bibfield  {author} {\bibinfo {author} {\bibfnamefont {D.}~\bibnamefont
  {Lee}}, \bibinfo {author} {\bibfnamefont {W.}~\bibnamefont {Jeong}}, \bibinfo
  {author} {\bibfnamefont {D.}~\bibnamefont {Yun}}, \bibinfo {author}
  {\bibfnamefont {S.-Y.}\ \bibnamefont {Park}}, \bibinfo {author}
  {\bibfnamefont {B.-K.}\ \bibnamefont {Ju}}, \bibinfo {author} {\bibfnamefont
  {K.-J.}\ \bibnamefont {Lee}}, \bibinfo {author} {\bibfnamefont {B.-C.}\
  \bibnamefont {Min}}, \bibinfo {author} {\bibfnamefont {H.~C.}\ \bibnamefont
  {Koo}},\ and\ \bibinfo {author} {\bibfnamefont {O.}~\bibnamefont {Lee}},\
  }\bibfield  {title} {\bibinfo {title} {Effects of interfacial oxidization on
  magnetic damping and spin–orbit torques},\ }\href
  {https://doi.org/10.1021/acsami.1c00608} {\bibfield  {journal} {\bibinfo
  {journal} {ACS Appl. Mater. Interfaces}\ }\textbf {\bibinfo {volume} {13}},\
  \bibinfo {pages} {19414} (\bibinfo {year} {2021})}\BibitemShut {NoStop}%
\bibitem [{\citenamefont {Fan}\ \emph {et~al.}(2020)\citenamefont {Fan},
  \citenamefont {Li}, \citenamefont {DC}, \citenamefont {Peterson},
  \citenamefont {Held}, \citenamefont {Sahu}, \citenamefont {Chen},
  \citenamefont {Zhang}, \citenamefont {Mkhoyan},\ and\ \citenamefont
  {Wang}}]{doi:10.1063/1.5124688YiHongWTe2}%
  \BibitemOpen
  \bibfield  {author} {\bibinfo {author} {\bibfnamefont {Y.}~\bibnamefont
  {Fan}}, \bibinfo {author} {\bibfnamefont {H.}~\bibnamefont {Li}}, \bibinfo
  {author} {\bibfnamefont {M.}~\bibnamefont {DC}}, \bibinfo {author}
  {\bibfnamefont {T.}~\bibnamefont {Peterson}}, \bibinfo {author}
  {\bibfnamefont {J.}~\bibnamefont {Held}}, \bibinfo {author} {\bibfnamefont
  {P.}~\bibnamefont {Sahu}}, \bibinfo {author} {\bibfnamefont {J.}~\bibnamefont
  {Chen}}, \bibinfo {author} {\bibfnamefont {D.}~\bibnamefont {Zhang}},
  \bibinfo {author} {\bibfnamefont {A.}~\bibnamefont {Mkhoyan}},\ and\ \bibinfo
  {author} {\bibfnamefont {J.-P.}\ \bibnamefont {Wang}},\ }\bibfield  {title}
  {\bibinfo {title} {Spin pumping and large field-like torque at room
  temperature in sputtered amorphous $\mathrm{WTe_{2-x}}$ films},\ }\href
  {https://doi.org/10.1063/1.5124688} {\bibfield  {journal} {\bibinfo
  {journal} {APL Mater.}\ }\textbf {\bibinfo {volume} {8}},\ \bibinfo {pages}
  {041102} (\bibinfo {year} {2020})}\BibitemShut {NoStop}%
\bibitem [{\citenamefont {DC}\ \emph {et~al.}(2018)\citenamefont {DC},
  \citenamefont {Grassi}, \citenamefont {Chen}, \citenamefont {Jamali},
  \citenamefont {Reifsnyder~Hickey}, \citenamefont {Zhang}, \citenamefont
  {Zhao}, \citenamefont {Li}, \citenamefont {Quarterman}, \citenamefont {Lv},
  \citenamefont {Li}, \citenamefont {Manchon}, \citenamefont {Mkhoyan},
  \citenamefont {Low},\ and\ \citenamefont {Wang}}]{DC2018}%
  \BibitemOpen
  \bibfield  {author} {\bibinfo {author} {\bibfnamefont {M.}~\bibnamefont
  {DC}}, \bibinfo {author} {\bibfnamefont {R.}~\bibnamefont {Grassi}}, \bibinfo
  {author} {\bibfnamefont {J.-Y.}\ \bibnamefont {Chen}}, \bibinfo {author}
  {\bibfnamefont {M.}~\bibnamefont {Jamali}}, \bibinfo {author} {\bibfnamefont
  {D.}~\bibnamefont {Reifsnyder~Hickey}}, \bibinfo {author} {\bibfnamefont
  {D.}~\bibnamefont {Zhang}}, \bibinfo {author} {\bibfnamefont
  {Z.}~\bibnamefont {Zhao}}, \bibinfo {author} {\bibfnamefont {H.}~\bibnamefont
  {Li}}, \bibinfo {author} {\bibfnamefont {P.}~\bibnamefont {Quarterman}},
  \bibinfo {author} {\bibfnamefont {Y.}~\bibnamefont {Lv}}, \bibinfo {author}
  {\bibfnamefont {M.}~\bibnamefont {Li}}, \bibinfo {author} {\bibfnamefont
  {A.}~\bibnamefont {Manchon}}, \bibinfo {author} {\bibfnamefont {K.~A.}\
  \bibnamefont {Mkhoyan}}, \bibinfo {author} {\bibfnamefont {T.}~\bibnamefont
  {Low}},\ and\ \bibinfo {author} {\bibfnamefont {J.-P.}\ \bibnamefont
  {Wang}},\ }\bibfield  {title} {\bibinfo {title} {Room-temperature high
  spin--orbit torque due to quantum confinement in sputtered
  $\mathrm{Bi_xSe_{(1-x)}}$ films},\ }\href
  {https://doi.org/10.1038/s41563-018-0136-z} {\bibfield  {journal} {\bibinfo
  {journal} {Nat. Mater.}\ }\textbf {\bibinfo {volume} {17}},\ \bibinfo {pages}
  {800} (\bibinfo {year} {2018})}\BibitemShut {NoStop}%
\end{thebibliography}
\providecommand{\noopsort}[1]{}\providecommand{\singleletter}[1]{#1}%

\end{document}